\documentclass[aps,pra,reprint,showpacs,notitlepage,superscriptaddress,twocolumn]{revtex4-2}
\usepackage{amssymb}
\usepackage{mathrsfs}
\usepackage{mathtools}
\usepackage{amsfonts}
\usepackage{graphicx}
\usepackage{amsmath}
\usepackage{dcolumn}
\usepackage{comment}
\usepackage{xcolor}
\usepackage{subfigure}
\usepackage{booktabs}
\usepackage{epsfig}
\usepackage{soul}
\usepackage[normalem]{ulem}
\usepackage[colorlinks,linkcolor=blue,citecolor=blue,hyperindex,bookmarks=false,pdfstartview=FitH,urlcolor=blue]{hyperref}

\providecommand{\U}[1]{\protect\rule{.1in}{.1in}}
\definecolor{Blue}{rgb}{0.00,0.00,0.80}
\definecolor{Red}{rgb}{0.80,0.00,0.00}
\newcommand{\blue}[1]{\textcolor{Blue}{#1}}

\newcommand{\figpanel}[2]{\hyperref[#1]{\ref*{#1}(#2)}}
\DeclarePairedDelimiter{\mean}{\langle}{\rangle}

\begin{document}

\title{Coherent feedback control for cavity optomechanical systems with a frequency-dependent mirror}

\author{Lei Du}
\affiliation{Department of Microtechnology and Nanoscience (MC2), Chalmers University of Technology, 412 96 Gothenburg, Sweden}
\author{Juliette Monsel}
\affiliation{Department of Microtechnology and Nanoscience (MC2), Chalmers University of Technology, 412 96 Gothenburg, Sweden}
\author{Witlef Wieczorek}
\affiliation{Department of Microtechnology and Nanoscience (MC2), Chalmers University of Technology, 412 96 Gothenburg, Sweden}
\author{Janine Splettstoesser}
\affiliation{Department of Microtechnology and Nanoscience (MC2), Chalmers University of Technology, 412 96 Gothenburg, Sweden}

\date{\today}
\begin{abstract}
Ground-state cooling of mechanical resonators is a prerequisite for the observation of various quantum effects in optomechanical systems and thus has always been a crucial task in quantum optomechanics.
In this paper, we study how to realize ground-state cooling of the mechanical mode in a Fano-mirror optomechanical setup, which allows for enhanced effective optomechanical interaction but typically works in the (deeply) unresolved-sideband regime.
We reveal that for such a two-sided cavity geometry with very different decay rates at the two cavity mirrors, it is possible to cool the mechanical mode down to its ground state within a broad range of parameters by using an appropriate single-sided coherent feedback. This is possible even if the total optical loss is more than seven orders of magnitude larger than the mechanical frequency and the feedback efficiency is relatively low.
Importantly, we show that a more standard double-sided feedback scheme is not appropriate to cooperate with a Fano-mirror system.
\end{abstract}

\maketitle

\section{Introduction}\label{secIntro}

Cavity optomechanics~\cite{OM2014RMP} provides an excellent platform for observing and harnessing quantum effects on a mesoscopic scale.
Optomechanical interactions, which typically arise from the momentum exchange between the electromagnetic field and a mechanical resonator, enable quantum control over photonic and phononic modes, leading to a series of important applications ranging from precise measurements~\cite{OMmeasure1,OMmeasure2,OMmeasure3}, quantum-state transfer~\cite{OMtransfer}, and frequency conversion~\cite{Fconversion1,Fconversion2,Fconversion3,Fconversion4}, to fundamental tests of quantum mechanics~\cite{Ftest1,Ftest2,Ftest3}.
In particular, cavity optomechanical systems can cool the mechanical degrees of freedom close to their quantum ground states, which is a key preliminary step towards witnessing genuine quantum phenomena~\cite{OMQT}.

\begin{figure}[bt]
\centering
\includegraphics[width=7.0 cm]{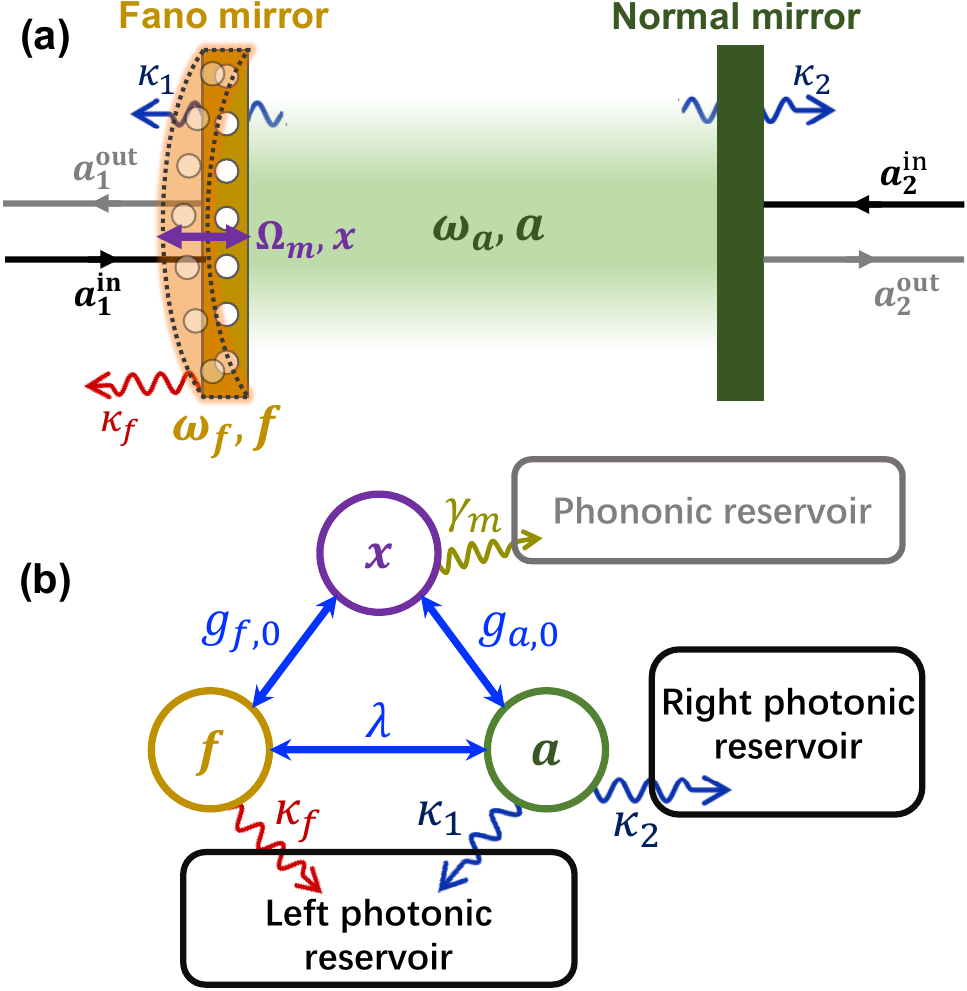}
\caption{(a) Schematics of the Fano-mirror optomechanical system. (b) Sketch of the interactions among the modes. The Fano mirror is a movable photonic crystal membrane that supports a mechanical vibration mode (with out-of-plane displacement $x$) as well as a guided optical mode (i.e., Fano mode) $f$. The mechanical and cavity modes are coupled through the radiation pressure, while the Fano and cavity modes are coupled through both the overlap of their electric fields and their couplings to the common (left) photonic reservoir. For some practical setups, the out-of-plane displacement of the membrane can also result in an in-plane mechanical strain, which induces a dispersive coupling between the Fano and mechanical modes.}\label{FanoModel}
\end{figure}

While the standard sideband cooling scheme~\cite{OM2014RMP} provides a powerful tool to access the mechanical ground state in optomechanical systems with high sideband resolution~\cite{MBF4,Teufel2011Jul, Delic2020Feb}, in practice setups often are in the unresolved-sideband regime, meaning that simple sideband cooling no longer works. Quantum feedback has emerged as a candidate for ground-state cooling and control of optomechanical systems that are in the unresolved-sideband regime~\cite{ZhangJreview}.
Compared to active feedback~\cite{MBF1,MBF2,MBF3,MBF5,MBF6,Magrini2021Jul,Tebbenjohanns2021Jul, Manikandan2023Feb}, that has to face excess noise in the out-of-loop optical field and decoherence due to quantum measurement~\cite{MBF6}, coherent feedback is promising since the quantum signals mediating the feedback can preserve their coherence~\cite{ZhangJreview,Lloyd2000}.
Coherent feedback has been suggested both to facilitate ground-state cooling in the resolved-sideband regime~\cite{LiJ2017,Harwood2021,HuangMPDI,FaradayR} and to allow ground-state cooling in the unresolved-sideband regime~\cite{JKGuo}, where we want to highlight a recent experimental realization~\cite{BaselPRX2023}. Ground-state cooling achieved by coherent feedback is, however, sensitive to limitations in the feedback efficiency, in particular for cavities with high losses.

A very different possibility to cool and control optomechanical systems that are originally in the unresolved-sideband regime is by introducing auxiliary quantum modes and engineering their interactions with the optomechanical system. It is thereby possible to create a narrow Fano resonance \cite{Limonov2017Sep}, with which the optomechanical system enters the resolved-sideband regime \textit{effectively}, and thus ground-state cooling is again allowed.
This strategy can be implemented, e.g., exploiting atomic ensembles~\cite{AE1,AE2,AE3} or double-cavity configurations~\cite{2cavity1,2cavity2,2cavity3,YCLiu2015pra}.
Most recently this mechanism has been demonstrated with even simpler geometric architectures, where two optical modes are coupled to each other and to a single mechanical mode~\cite{GenesDoped2014,FanoM,GenesQST,Juliette2021pra,WWoe2023}, see also Fig.~\ref{FanoModel}.
In particular, recent experiments~\cite{WWoe2023,Mitra2024Apr} demonstrated a linewidth reduction of the optomechanical cavity by using a frequency-dependent photonic-crystal membrane mirror, where the guided optical mode plays the role of the Fano mode.  In these two cases, the underlying system was a microcavity. The motivation for realizing such systems is that the single-photon optomechanical coupling is increased with respect to typical optomechanical cavities, with the prospect of even accessing the ultrastrong coupling regime for certain system parameters. However, these systems have so far the drawback that they are, despite their linewidth reduction, still far from the resolved-sideband regime.
Though, in theory, the effective optical losses could be further reduced and made smaller than the mechanical frequency~\cite{Juliette2023arxiv}, it still remains a challenge to obtain ideal parameters for mode-coupling and Fano resonance, where the quantum regime~\cite{SVCR1, SVCR2, Wise2024May} can actually be reached.

In this paper, we study how to facilitate ground-state cooling of the mechanical mode in such a Fano-mirror optomechanical system by combining it with a realistic coherent feedback scheme.
While for such a two-sided standing-wave quantum system, a double-sided coherent feedback scheme might seem a good choice~\cite{LiJ2017,HuangMPDI,FaradayR}, we here show that it becomes inappropriate when the cavity has very different decay rates at the two end mirrors, as it is the case for microcavities with a movable Fano mirror, like in Fig.~\ref{FanoModel}.
Instead, we reveal that a suitable single-sided coherent feedback behaves as the ideal candidate, despite the fact that its efficiency is inherently low in concrete realizations.
Indeed, we demonstrate that this combination of single-sided coherent feedback and Fano mode enables ground-state cooling in a \textit{deeply} unresolved sideband regime where the total cavity decay rate is about seven orders of magnitude larger than the mechanical frequency. Such an achievement would be impossible, within a wide range of realistic parameters, based on \textit{only} the Fano mode or \textit{only} the coherent feedback. Moreover, considering that some relevant parameters, such as the coherent coupling strength between the cavity and Fano modes, are challenging to control precisely in experiments, the coherent feedback provides a controllable knob with which ground-state cooling is still allowed even if the actual parameters deviate from the desired values.

The remainder of this paper is organized as follows. We first describe the model for an optomechanical cavity with a Fano mirror in Sec.~\ref{SecModel}. Then, in Sec.~\ref{secDb}, we address the general effect of two-sided coherent feedback and demonstrate that it is inappropriate for the Fano-mirror setup of Sec.~\ref{SecModel}. Finally, Sec.~\ref{secSg} shows how ground-state cooling can instead be achieved for realistic parameters, if the Fano-mirror setup is combined with a \emph{single-sided} coherent feedback scheme. Technical details are presented in the Appendix. Throughout the manuscript, we take $\hbar=1$.

\section{Cavity optomechanics with a Fano mirror}\label{SecModel}

We consider a standing-wave optomechanical system, with a Fabry-Pérot-type geometry, but where one of the cavity mirrors exhibits a strongly frequency-dependent response.
Such Fano mirrors enable normal (hybrid) modes with a normal-mode linewidth which is drastically decreased with respect to the bare linewidth of the optical modes~\cite{FanoM,WWoe2023,Juliette2023arxiv,Mitra2024Apr}.

Concretely, as shown in Fig.~\figpanel{FanoModel}{a}, we have a standard cavity mode $a$, with frequency $\omega_{a}$. The left cavity mirror supports both a mechanical vibrational mode, with out-of-plane dimensionless displacement $x$ and frequency $\Omega_{m}$, and a guided optical mode $f$ with frequency $\omega_{f}$. We refer to the latter as \textit{Fano mode}. It can be experimentally realized by a suspended dielectric membrane with designed subwavelength photonic crystal structures~\cite{WWoe2023}.
As depicted in Fig.~\figpanel{FanoModel}{b}, the cavity mode is coupled to the mechanical mode via radiation pressure, resulting in the single-photon coupling strength $g_{a,0}$. Meanwhile, the Fano mode also couples to the mechanical mode: in many practical setups, the out-of-plane displacement of the membrane will also result in an in-plane mechanical strain, which alters the optical properties of the membrane and thus induces a dispersive (i.e., radiation-pressure like) coupling between the Fano and mechanical modes~\cite{WWoe2023}. We denote the associated single-photon coupling strength $g_{f,0}$.
In addition, the cavity mode coherently interacts with the Fano mode due to the overlap of their electric fields, with coupling strength $\lambda$~\cite{FanoM,GenesQST}.
As a result, the Hamiltonian of such a Fano-mirror optomechanical system
can then be written as
\begin{equation}
\begin{split}
H&=\omega_{a}a^{\dagger}a+\omega_{f}f^{\dagger}f+\frac{\Omega_{m}}{2}\left(x^{2}+p^{2}\right) \\
&\quad\,+\left(g_{a,0}a^{\dagger}a+g_{f,0}f^{\dagger}f\right)x+\lambda(a^{\dagger}f+f^\dagger a).
\end{split}
\label{totalH}
\end{equation}
The (dimensionless) momentum operator of the mechanical mode $p$ satisfies the commutation relation $[x, p]=i$.

Furthermore, the cavity is a two-sided system, which is coupled to an electromagnetic environment on each side, with typically very different decay rates $\kappa_1$ and $\kappa_2$ at the left and right mirrors, respectively~\cite{WWoe2023}. The right normal mirror, which is frequency-independent and, e.g., realized by a distributed Bragg reflector, has, in contrast to the left (movable, Fano) mirror, very low transmissivity~\cite{FanoM}. The Fano mode is also coupled to the left electromagnetic environment [see Fig.~\figpanel{FanoModel}{b}], with loss rate $\kappa_f$. Both optical modes are coupled to the same environment, which gives rise to a dissipative coupling of strength $\kappa_{1f}=\sqrt{\kappa_{1}\kappa_{f}}$~\cite{FanoM}. The overall coupling between the cavity mode and Fano mode is thus $G=\lambda-i\kappa_{1f}$.

Strictly speaking, for microcavities there are also dissipative optomechanical couplings which arise from the dependence of the optical decay rates on the mechanical displacement. Moreover, the mechanical displacement can also modulate the coherent coupling $\lambda$ between the cavity and Fano modes, which, in the linearized Hamiltonian, is equivalent to introducing finite shifts to the couplings among the optical and mechanical modes~\cite{GenesQST}. However, in the case considered in this paper--where the cavity length is sufficiently large so that the Fano mirror does not ``feel'' the optical field of the other mirror--both the dissipative optomechanical couplings and the position dependence of $\lambda$ can be safely neglected. In particular, recent studies have shown that, with the parameters considered below, the contributions of the dissipative optomechanical couplings (for both the cavity and Fano modes) can be negligible~\cite{Juliette2023arxiv}. Given these features, we consider in this paper relatively weak single-photon optomechanical coupling $g_{a,0}$.

We now assume that the system is driven by a coherent pumping field with amplitude $\varepsilon_\mathrm{p}$ and frequency $\omega_\mathrm{p}$. The pumping amplitude $\varepsilon_{\mathrm{p}}=\sqrt{2\kappa_{1(2)}\mathcal{P}/(\hbar\omega_\mathrm{p})}$ (assumed to be real without loss of generality) is related to the power $\mathcal{P}$ of the pumping field, with $\kappa_{1}$ or $\kappa_{2}$ chosen depending on which mirror the pumping field is applied on.
In this section, we do not need to specify which mirror the pumping field is applied on, since this only affects the steady-state values of the two optical modes. If the pumping is strong, the dynamics of the system can be linearized~\cite{OM2014RMP}. We therefore write each operator $o$ as the sum of its steady-state mean value $\bar{o}$~\footnote{In order to access well-defined steady-state mean values, we examine the stability of the system by using the Routh-Hurwitz criterion~\cite{DeJesus1987Jun}. We indicate the unstable regime by white areas in all density plots of this paper.} and a quantum fluctuation $\delta o$, i.e., $o=\bar{o}+\delta o$. The linearized quantum Langevin equations of the whole system are then~\cite{Juliette2023arxiv}
\begin{subequations}
    \begin{eqnarray}
\delta\dot{x}&=&\Omega_{m}\delta p, \label{xFano} \\
\delta\dot{p}&=&-\Omega_{m}\delta x-\gamma_{m}\delta p-g_{a}^{*}\delta a-g_{a}\delta a^{\dag} \nonumber\\
&&-g_{f}^{*}\delta f-g_{f}\delta f^{\dag}+\sqrt{2\gamma_{m}}\xi_{m}, \label{pFano} \\
\delta\dot{a}&=&-\left(i\Delta_{a}+\kappa_\mathrm{tot}\right)\delta a-ig_{a}\delta x \nonumber\\
&&-iG\delta f+\sum_{j=1,2}\sqrt{2\kappa_{j}}a_{j}^{\text{in}}, \label{aFano}\\
\delta\dot{f}&=&-\left(i\Delta_{f}+\kappa_{f}\right)\delta f-ig_{f}\delta x \nonumber\\
&&-iG\delta a+\sqrt{2\kappa_{f}}a_{1}^{\text{in}}. \label{fFano}
\end{eqnarray}
\end{subequations}
Here, $g_{a}=g_{a,0}\bar{a}$  describes the enhanced optomechanical interaction between the mechanical and cavity modes and $g_{f}=g_{f,0}\bar{f}$ the one between the mechanical and Fano modes.
The frequency detunings between the pumping field and the two optical 
modes are given by $\Delta_{a}=\omega_{a}-\omega_{p}+g_{a,0}\bar{x}$ (for the cavity mode) and by $\Delta_{f}=\omega_{f}-\omega_{p}+g_{f,0}\bar{x}$ (for the Fano mode), including the influence of the steady-state mechanical displacement.
Furthermore, the decay rate of the cavity mode to the right photonic reservoir (through the normal mirror), $\kappa_{2}$, enters these equations---also through the total decay rate $\kappa_\mathrm{tot}=\kappa_1+\kappa_2$. The damping rate of the mechanical mode is $\gamma_{m}$.
Furthermore, $a_{1}^{\text{in}}$ and $a_{2}^{\text{in}}$ are the vacuum input noises from the left and right environments, respectively, which have the only nonvanishing correlation function $\mean{a_{i}^{\text{in}}(t)(a_{i}^{\text{in}})^\dagger(t')} = \delta(t-t')$, with $i = 1,2$.
$\xi_{m}$ is the Brownian thermal noise of the mechanical resonator, which satisfies $\langle\xi_{m}(t)\xi_{m}(t')+\xi_{m}(t')\xi_{m}(t)\rangle\approx(n_{m}+1/2)\delta(t-t')$ with $n_{m}$ the thermal phonon number~\footnote{This approximation is known to give a spurious term in the phase fluctuation spectrum \cite{BrownianN}, but it does not impact the quantity we are interested in, namely the steady-state phonon number, as discussed, e.g., in Appendix of a previous work \cite{Juliette2021pra}}.
We here assume complex optomechanical coupling coefficients, since in general $\bar{a}$ and $\bar{f}$ are not real simultaneously, as will be shown in Sec.~\ref{secSgb}.

As mentioned, the most intriguing feature of the optomechanical system with a Fano mirror is the presence of a normal-mode optical resonance which has a very narrow linewidth, compared with the original linewidth of both the pure cavity and the Fano mode. This important property is the basis for achieving mechanical ground-state cooling, for a system which would otherwise be in the unresolved-sideband regime, namely with a cavity linewidth that is larger than the mechanical frequency~\cite{Juliette2021pra,Juliette2023arxiv}.
The narrow Fano resonance can be understood from the \textit{normal modes} of the cavity and Fano modes where, in order to clarify this concept, we drop for the moment the optomechanical interactions. For the model in Fig.~\ref{FanoModel}, the complex energies of the two normal modes can then be written as~\cite{WWoe2023,Juliette2023arxiv}
\begin{equation}
\begin{split}
\tilde{\omega}_{\pm}&=\frac{\Delta_{a}+\Delta_{f}}{2}-i\frac{\kappa_\mathrm{tot}+\kappa_{f}}{2}\\
&\quad\,\pm\sqrt{\left(\frac{\Delta_{a}-\Delta_{f}}{2}-i\frac{\kappa_\mathrm{tot}-\kappa_{f}}{2}\right)^{2}+G^{2}},
\end{split}
\label{normalmodes0}
\end{equation}
which corresponds to normal-mode resonance frequencies $\omega_{\pm}=\text{Re}(\tilde{\omega}_{\pm})$ and normal-mode linewidths $\kappa_{\pm}=-\text{Im}(\tilde{\omega}_{\pm})$.
With appropriate parameters, one of the normal modes can have a linewidth that is several orders of magnitude smaller than $\kappa_{\text{tot}}$ and $\kappa_{f}$ (and even smaller than commonly realized mechanical frequencies), thus rendering the originally unresolved-sideband optomechanical system sideband-resolved~\cite{FanoM,Juliette2023arxiv}.

While the above Fano resonance provides an opportunity for ground-state cooling for optomechanical setups that are originally in the highly unresolved-sideband regime, in experiments, it is challenging to meet the required parametric conditions, see Ref.~\cite{Juliette2023arxiv} for more details about relevant parameter regimes. In particular, some relevant parameters, such as the coherent coupling strength between the cavity and Fano modes, are difficult to precisely engineer and control. In this paper, we aim to design a feasible coherent feedback scheme, with which the cooling effect of the Fano-mirror optomechanical system can be enabled and enhanced across a broad range of parameters. As will be shown below, thanks to the interplay between the Fano resonance and an appropriate coherent feedback, ground-state cooling can be realized even if the system is originally in the deeply unresolved-sideband regime (with the total decay rate of the cavity mode more than seven orders of magnitude larger than the mechanical frequency) and if the Fano-mirror parameters alone are not ideal. This is possible even if the efficiency of the coherent feedback loop is low.

\section{Double-sided coherent feedback}\label{secDb}

When dealing with a two-port quantum system, such as the optomechanical cavity introduced above, one can construct a double-sided coherent feedback where the output field from one port is fed back to the other via a (unidirectional) traveling-wave field~\cite{ZhangJreview}. This double-sided feedback scheme has the advantage that the feedback efficiency can be in principle very large---even close to unity. This will become more apparent when comparing it to the single-sided feedback scheme considered in Sec.~\ref{secSg}.

However, we show in this section that such a feedback scheme works well only for $\kappa_{1} \simeq \kappa_2$ by analyzing, for simplicity, the case of a standard Fabry-Pérot cavity and conclude that it is therefore not suitable for the system described in Sec.~\ref{SecModel} since the Fano mirror makes the cavity highly asymmetric, with $\kappa_1 \gg \kappa_2$.

\subsection{Equations of motion with feedback}\label{secDba}

\begin{figure}[ptb]
\centering
\includegraphics[width=6.8 cm]{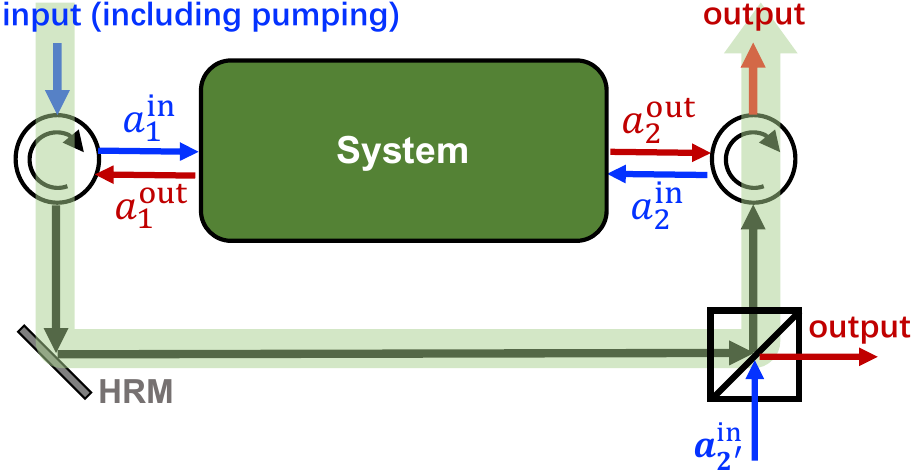}
\caption{Schematics of a possible implementation of the considered double-sided feedback loop, which includes highly reflected mirrors (HRMs) and circulators. The faint green arrow indicates the traveling-wave character of the feedback, going along with a phase accumulation $\phi$. To include the practically unavoidable loss and imperfection of the feedback loop, we also introduce an additional beam splitter, where an extra input noise is injected. The system can be an arbitrary two-port quantum system, such as the two-sided optomechanical cavity considered throughout this paper.}\label{Dbloop}
\end{figure}

We consider a two-sided Fabry-Perot-type optomechanical system, where the two cavity mirrors serve as the two ports---indicated by a green box in Fig.~\ref{Dbloop}. The feedback loop is implemented with circulators and highly-reflecting mirrors (HRMs) as shown in Fig.~\ref{Dbloop} (see Appendix~\ref{appSL} for more details about possible experimental implementations). Then, the linearized quantum Langevin equation of the cavity mode $a$ (in the rotating frame with respect to the driving frequency) is given by
\begin{equation}
\begin{split}
\delta\dot{a}&=-(i\Delta_{a}+\kappa_\mathrm{tot})\delta a-ig_{a}\delta x\\
&\quad\,+\sqrt{2\kappa_{1}}a_{1}^{\text{in}}+\sqrt{2\kappa_{2}}a_{2}^{\text{in}}.
\end{split}
\label{Dba}
\end{equation}
Here, all the symbols have the same meaning as in the last section. We consider an optomechanical system with standard mirrors, namely no Fano mode is present. We furthermore assume that the mechanical mode is placed inside the cavity (for example via a standard membrane-in-the-middle setup) such that the coupling to the coherent feedback is not impacted by the mechanical motion.

In order to introduce the effect of the coherent feedback, we now plug in the input-output relations
\begin{subequations}
\begin{eqnarray}
    a_{2}^{\text{out}} &= &\sqrt{2\kappa_{2}}a-a_{2}^{\text{in}}\\
    a_{2}^{\text{in}}&=&a_{1}^{\text{out}}\sqrt{\eta}\text{exp}(i\phi)+\sqrt{1-\eta}a_{2'}^{\text{in}}\\
    a_{1}^{\text{out}} & = & \sqrt{2\kappa_{1}}a-a_{1}^{\text{in}}
\end{eqnarray}
\end{subequations}
into Eq.~(\ref{Dba}). Here, $0\leq\eta<1$ is the efficiency of the feedback loop, which can be smaller than one due to unavoidable losses and imperfections. The extra input noise, which arises from the practically unavoidable loss and imperfection of the feedback loop, is described by $a_{2'}^{\text{in}}$ (in Fig.~\ref{Dbloop} it comes from the beam splitter). Furthermore, $\phi$ is the phase accumulation of the field traveling from one mirror to the other. With this, we have
\begin{equation}
\begin{split}
\delta\dot{a}&=-\left(i\Delta_\mathrm{eff}+\kappa_\mathrm{tot,eff}\right)\delta a-ig_{a}\delta x\\
&\quad\,+\left(\sqrt{2\kappa_{1}}-\sqrt{2\kappa_{2}\eta} e^{i\phi}\right)a_{1}^{\text{in}}+\sqrt{2\kappa_{2}(1-\eta)}a_{2'}^{\text{in}},
\end{split}
\label{Dba2}
\end{equation}
where the effective parameters result from the coherent feedback.
Specifically, we have introduced the effective total decay rate of the cavity mode, $\kappa_{\text{tot,eff}}=\kappa_\mathrm{tot}-2\kappa_{12}\sqrt{\eta}\cos{\phi}$ with  $\kappa_{12}=\sqrt{\kappa_{1}\kappa_{2}}$,  which can be reduced with respect to the original decay rates, depending on the efficiency $\eta$, as well as the phase accumulation $\phi$. Moreover, the input noise of the cavity is suppressed in a similar, but importantly not identical manner.
The coupling between cavity mode and environment, e.g., $\sqrt{2\kappa_{1}}-\sqrt{2\kappa_{2}\eta} e^{i\phi}$, can be even complex, containing both coherent and dissipative components.
Such a double-sided coherent feedback can thus potentially provide the possibility to realize mechanical ground-state cooling in the \textit{originally} sideband-unresolved regime. In addition, the coherent feedback also modifies the resonance frequency of the cavity mode such that the effective detuning is given by $\Delta_{\text{eff}}=\Delta_{a}-2\kappa_{12}\sqrt{\eta}\sin{\phi}$.

If the pumping field is also applied through the traveling-wave path, the steady-state mean value of the cavity mode can be obtained as $\bar{a}=\varepsilon_\mathrm{p,eff}/(i\Delta_\mathrm{eff}+\kappa_\mathrm{tot,eff})$ with $\varepsilon_\mathrm{p,eff}=(\sqrt{2\kappa_{1}}-\sqrt{2\kappa_{2}\eta} e^{i\phi})\sqrt{\mathcal{P}/(\hbar\omega_\mathrm{p}})\text{exp}(i\theta)$. Here $\theta$ is the phase of the pumping field, which can be readily controlled in experiments. In view of this, $\varepsilon_\mathrm{p,eff}$ can be tuned to be real by appropriately choosing $\theta$.

\subsection{Feedback-assisted ground-state cooling}\label{secDbb}
From the linearized equations of motion of the optomechanical system, we can compute the steady-state phonon number of the mechanical mode,
\begin{equation}
    n_\mathrm{fin}=\frac{1}{2}(\langle\delta x^{2}\rangle+\langle\delta p^{2}\rangle-1),\label{phononN}
\end{equation}
by solving the Lyapunov equation describing the evolution of the second-order moments of the system, see details in Appendix~\ref{appLyap}.

\begin{figure}[ptb]
\centering
\includegraphics[width=1.0\linewidth]{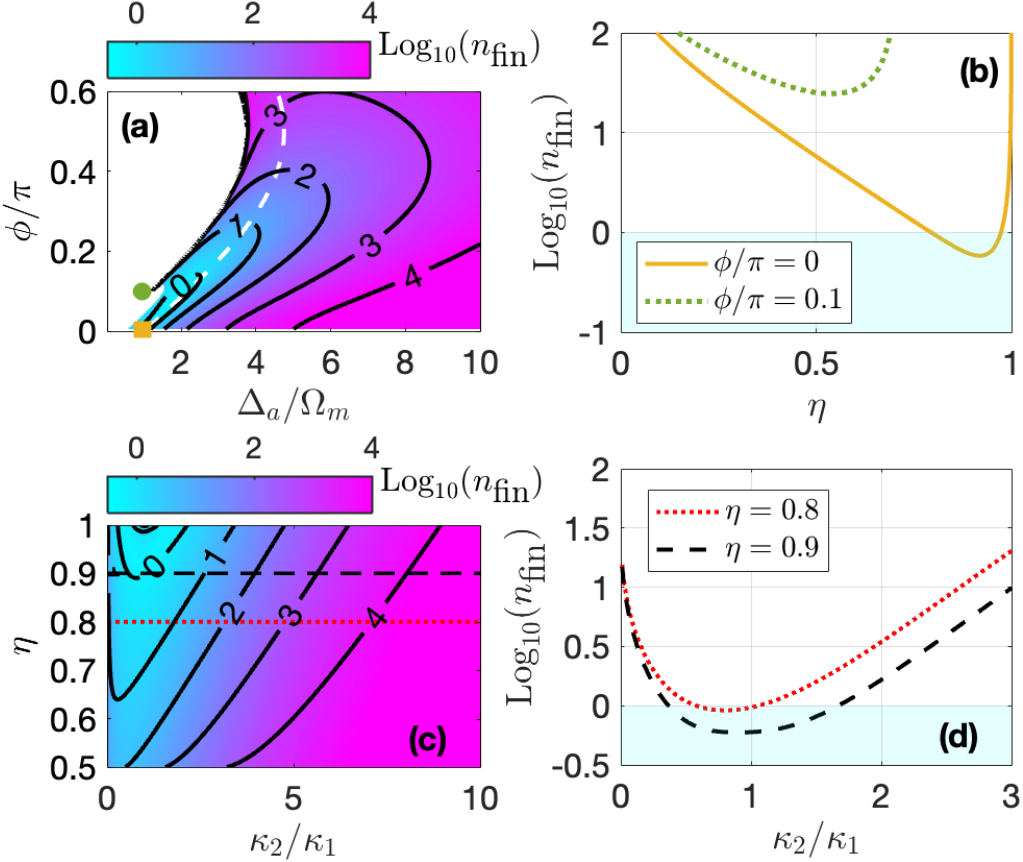}
\caption{Plots of the final phonon number $n_\mathrm{fin}$ (on logarithmic scale) for the double-sided feedback scheme. (a) Final phonon number versus detuning $\Delta_{a}$ and phase accumulation $\phi$ with $\eta=0.95$ and $\kappa_{2}=\kappa_{1}$. (b) Final phonon number versus feedback efficiency $\eta$ with $\Delta_{a}/\Omega_{m}=1$, $\kappa_{2}=\kappa_{1}$, and two selected values of $\phi$ [with the orange solid and green dotted lines corresponding to the orange square and green circle in (a), respectively]. The orange curve approaches the minimum (i.e., $n_{\text{fin}}\approx0.59$) for $\eta\approx0.92$. (c) Final phonon number versus decay ratio $\kappa_{2}/\kappa_{1}$ and feedback efficiency $\eta$ with $\Delta_{a}/\Omega_{m}=1$ and $\phi=0$. (d) Final phonon number versus decay ratio $\kappa_{2}/\kappa_{1}$ with $\Delta_{a}/\Omega_{m}=1$, $\phi=0$, and two selected values of $\eta$ [with the black dashed and red dotted lines corresponding to the same types of line in (c), respectively.] The white dashed line in (a) indicates $\Delta_{\text{eff}}=\Omega_{m}$ while the cyan areas in (b) and (d) represent the regimes of $n_\mathrm{fin}<1$. Other parameters are $\kappa_{1}/2\pi=0.25\,\text{MHz}$, $\Omega_{m}/2\pi=0.13\,\text{MHz}$, $\gamma_{m}/2\pi=0.12\,\text{Hz}$, $g_{0}/2\pi=50\,\text{Hz}$, $|\varepsilon_\mathrm{p,eff}|/2\pi=80\,\text{MHz}$, and $\bar{n}_{m}=9.6\times 10^{4}$~\cite{Harris2008nature}.}
\label{DbNf}
\end{figure}

In Fig.~\ref{DbNf}, we provide a proof-of-principle demonstration for mechanical ground-state cooling with the help of double-sided coherent feedback.
Here---and also in the following figures of this paper---we always test the stability of the system, show results for the stable regime, and leave the unstable regime white. We adopt a set of experimentally available parameters as in Ref.~\cite{Harris2008nature}, which implies that the system is originally in the unresolved-sideband regime.
It is clear from Figs.~\figpanel{DbNf}{a} and \figpanel{DbNf}{b} that ground-state cooling (i.e., $n_\mathrm{fin}<1$) can be achieved in the presence of the feedback.
Note that an extremely small $\phi$ (especially $\phi=0$) may not be a physically feasible choice for the implementation in Fig.~\ref{Dbloop}. However, both $\Delta_{\text{eff}}$ and $\kappa_{\text{tot},\text{eff}}$ exhibit a phase dependence with a periodicity of $2\pi$. In view of this, the phase accumulation $\phi$ can be interpreted as $\text{mod}(\phi,2\pi)$.
The minimum of the final phonon number is always located at the \textit{effective mechanical red sideband}, namely at the phase-dependent detuning $\Delta_{\text{eff}}=\Omega_{m}$, as shown by the white dashed line in Fig.~\figpanel{DbNf}{a}.
Note that genuine ground-state cooling also demands the equipartition of energy, namely $\langle\delta x^{2}\rangle\simeq\langle\delta p^{2}\rangle$.
This criterion is also checked as shown in Fig.~\ref{Equip} in Appendix~\ref{appEqp}.

Figure~\figpanel{DbNf}{b} shows that increasing the feedback efficiency further enhances the cooling effect, but only up to a point. For higher values of $\eta$,  the cooling effect is weakened again and the system then enters the unstable regime.
This is consistent with the fact that, in sideband cooling schemes, the decay of the cavity field should be sufficiently strong to allow the energy to flow from the mechanical mode to the cavity mode and then to the environment.

We would like to point out that the effect of the double-sided coherent feedback is strongly dependent on the \textit{geometric symmetry} of the cavity, i.e., the ratio $\kappa_{2}/\kappa_{1}$ between the decay rates at the two mirrors.
As shown in Fig.~\figpanel{DbNf}{c}, the minimum of $n_\mathrm{fin}$ always appears when $\kappa_{2}/\kappa_{1}\approx1$. This can be understood again from the effective decay rate $\kappa_{\text{tot,eff}}$ of the cavity, which is significantly reduced when $\kappa_{1}\approx\kappa_{2}$, $\text{mod}(\phi, 2\pi)=0$, and $\eta\rightarrow1$.
Moreover, we focus in Fig.~\figpanel{DbNf}{d} on two selected cases ($\eta=0.8$ and $0.9$) where the system is always stable when $\kappa_{2}/\kappa_{1}$ varies. One can find that $n_\mathrm{fin}$ increases rapidly as the system deviates from the condition $\kappa_{2}/\kappa_{1}\sim1$, showing how this asymmetry hinders ground-state cooling.
Even with identical mirrors (i.e., $\kappa_{1}=\kappa_{2}$), the coherent feedback is further limited by its efficiency, especially for bad cavities. Specifically, to reach an effective sideband-resolved regime, the feedback efficiency must satisfy $1-\sqrt{\eta}<\Omega_{m}/\kappa_{\text{tot}}$, which is a stringent requirement for bad cavities (e.g., $\kappa_{\text{tot}}/\Omega_{m}\sim10^{7}$ as will be considered in Sec.~\ref{secSg}). 
In view of this, the double-sided coherent feedback scheme is not a suitable candidate to facilitate ground-state cooling in either the aforementioned Fano-mirror optomechanical setup, which \textit{by construction} shows a highly asymmetric two-sided geometry, or conventional optomechanical setups with a poor sideband resolution $\Omega_{m}/\kappa_{\text{tot}}$. This conclusion is further verified in Appendix~\ref{appNormal}, where we show that for $\kappa_{2}\ll\kappa_{1}$ such a double-sided coherent feedback can hardly affect the two optical normal modes.

\begin{figure}[ptb]
\centering
\includegraphics[width=8.5 cm]{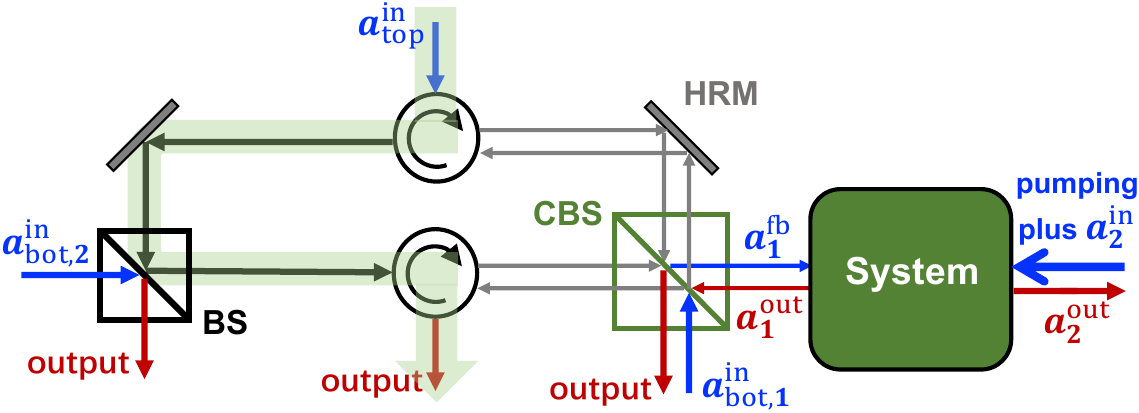}
\caption{Schematics of a possible implementation of the considered single-sided feedback loop, including HRMs, circulators, and (controllable) beam splitters (BS and CBS). The faint green arrow indicates the traveling-wave character of the feedback.  The system can be an arbitrary two-port quantum system, such as a two-sided optomechanical cavity considered throughout this paper. The feedback scheme continues being effective even for highly asymmetric cavities.}\label{Sloop}
\end{figure}

\section{Single-sided coherent feedback}\label{secSg}

 In Sec.~\ref{secDb}, we have shown that the double-sided feedback loop in Fig.~\ref{Dbloop} becomes inefficient if the cavity has very different decay rates at the two mirrors, which is typically true for the Fano-mirror optomechanical system considered in this paper. In view of this, we now consider a \textit{single-sided} coherent feedback scheme as shown in Fig.~\ref{Sloop}. The output field from the Fano mirror (which typically has a much larger decay rate than the right normal mirror) will be fed back to the Fano mirror again. Such a possible implementation also involves circulators and HRMs but arranged differently compared to the double-sided feedback scheme (see Appendix~\ref{appSL} for more details). Furthermore, in order to couple the fields of two different circulators to the same cavity mode through only one port (mirror), we also exploit a controllable beam splitter (CBS) with transmission coefficient $t_\mathrm{CBS}$ and reflection coefficient $r_\mathrm{CBS}$. As in the double-sided feedback scheme, an additional beam splitter with reflection coefficient $r_{\text{ex}}=\sqrt{\eta_{\text{ex}}}$ is introduced to account for the practically unavoidable loss and imperfection in the feedback loop. At first glance, such an implementation comes at the cost of a reduced feedback efficiency, which, in the ideal limit of $\eta_{\text{ex}}=1$, is determined by $t_\mathrm{CBS}$ and $r_\mathrm{CBS}$ as will be shown below. This is not true, however, since one can still expect a perfect destructive interference (the external decay of the two optical modes through the Fano mirror can be totally suppressed, as shown by their effective decay rates derived below). Moreover, we reveal that the cooling effect can be significantly enhanced even when $\eta_{\text{ex}}<1$.
\subsection{Equations of motion with feedback}\label{secSga}

In order to write down the modified Langevin equations, we first note that, in this case, the whole input field $a_{1}^{\text{in}}$ upon the left Fano mirror---before being modified by the coherent feedback---results from a superposition of three parts, which are described by $a_{\text{top}}^{\text{in}}$, $a_{\text{bot},1}^{\text{in}}$, and $a_{\text{bot},2}^{\text{in}}$~\footnote{Actually, the input noise $a_{\text{bot},1}^{\text{in}}$ is the superposition of both the input field coming from the ``bottom environment'' and the output field coming from the Fano mirror, so that it also satisfies the canonical commutation relation.}, as shown in Fig.~\ref{Sloop}.
These three parts eventually combine at the controllable beam splitter and thus one has
\begin{align}
a_{1}^{\text{in}}&=r_\mathrm{CBS}a_\mathrm{top}^{\text{in}}+t_\mathrm{CBS}(\sqrt{\eta_{\text{ex}}}a_\mathrm{bot,1}^{\text{in}}+\sqrt{1-\eta_{\text{ex}}}a_\mathrm{bot,2}^{\text{in}})\nonumber \\
&=r_\mathrm{CBS}a_\mathrm{top}^{\text{in}}+t_\mathrm{CBS}a_\mathrm{bot}^{\text{in}}.
\label{originalin}
\end{align}
In the above definition, we have (i) absorbed the trivial phases into the input noise operators $a_\mathrm{top}^{\text{in}}=(X_\mathrm{top}^{\text{in}}+iP_\mathrm{top}^{\text{in}})/\sqrt{2}$ and $a_\mathrm{bot,j}^{\text{in}}=(X_\mathrm{bot,j}^{\text{in}}+iP_\mathrm{bot,j}^{\text{in}})/\sqrt{2}$ ($j=1,2$) and (ii) defined a new input noise operator $a_{\text{bot}}^{\text{in}}=\sqrt{\eta_{\text{ex}}}a_\mathrm{bot,1}^{\text{in}}+\sqrt{1-\eta_{\text{ex}}}a_\mathrm{bot,2}^{\text{in}}$. Note that all the input noise operators, including their combinations $a_{\text{bot}}^{\text{in}}$ and $a_{1}^{\text{in}}$, satisfy the canonical commutation relation $[a_{j}^{\text{in}}(t), a_{j'}^{\text{in}}(t')^{\dagger}]=\delta_{j,j'}\delta(t-t')$ and the correlation function is $\langle a_{j}^{\text{in}}(t)a_{j'}^{\text{in}}(t')^{\dagger}\rangle=\delta_{j,j'}\delta(t-t')$.
Moreover, the coherent feedback is mediated by the traveling field experiencing a reflection and a transmission at the beam splitter. The actual input field at the Fano mirror, including feedback, is hence given by
\begin{equation}
\begin{split}
a_{1}^{\text{fb}}&=\sqrt{\eta}e^{i\phi}a_{1}^{\text{out}}+a_{1}^{\text{in}} \\
&=\sqrt{\eta}e^{i\phi}\left(\sqrt{2\kappa_{1}}a+\sqrt{2\kappa_{f}}f-a_{1}^{\text{in}}\right)+a_{1}^{\text{in}} \\
&=\sqrt{\eta}e^{i\phi}\left(\sqrt{2\kappa_{1}}a+\sqrt{2\kappa_{f}}f\right) \\
&\quad\,+\left(1-\sqrt{\eta}e^{i\phi}\right)\left(r_\mathrm{CBS}a_\mathrm{top}^{\text{in}}+t_\mathrm{CBS}a_\mathrm{bot}^{\text{in}}\right),
\end{split}
\label{Sloopfb}
\end{equation}
where $\phi$ again includes all possible phase shifts, such as the phase difference between the reflection and transmission fields of the beam splitters. The overall efficiency of the feedback loop is then defined as $\eta=t_\mathrm{CBS}^{2}r_\mathrm{CBS}^{2}\eta_{\text{ex}}$, including the contribution of the additional loss. The input-output relation
\begin{equation}
    a_{2}^{\text{out}}=\sqrt{2\kappa_{2}}a-a_{2}^{\text{in}}
\end{equation}
at the right normal mirror is identical to that of a conventional setup. Substituting Eq.~(\ref{Sloopfb}) into Eqs.~(\ref{aFano}) and (\ref{fFano}), we have
\begin{subequations}\label{eq:FanoLangevin}
\begin{eqnarray}
\delta\dot{a}&=&-\left[i\Delta_{a,\mathrm{eff}}+\kappa_{a,\mathrm{eff}}\right]\delta a -ig_{a}\delta x-iG_\mathrm{eff}\delta f \nonumber\\
&&+\sqrt{2\kappa_{1}}\left(1-\sqrt{\eta}e^{i\phi}\right)a_{1}^{\text{in}}+\sqrt{2\kappa_{2}}a_{2}^{\text{in}}, \label{aFanoSF}\\
\delta\dot{f}&=&-\left[i\Delta_{f,\mathrm{eff}}+\kappa_{f,\mathrm{eff}}\right]\delta f -ig_{f}\delta x-iG_\mathrm{eff}\delta a \nonumber\\
&&+\sqrt{2\kappa_{f}}\left(1-\sqrt{\eta}e^{i\phi}\right)a_{1}^{\text{in}}.\label{fFanoSF}
\end{eqnarray}
\end{subequations}
Clearly, such a single-sided coherent feedback modifies not only the decay rates and the corresponding input noises of the two optical modes, but also their dissipative coupling through the left photonic reservoir.
While the maximum efficiency is $\eta = 0.25$ by definition, the cooling effect can still be greatly enhanced with the combination of the coherent feedback and Fano resonance, as will be shown below.
Concretely, $\Delta_{a,\text{eff}}=\Delta_{a}-2\kappa_{1}\sqrt{\eta}\sin{\phi}$ and $\Delta_{f,\text{eff}}=\Delta_{f}-2\kappa_{f}\sqrt{\eta}\sin{\phi}$ are the effective detunings of the cavity and Fano modes; $\kappa_{a,\text{eff}}=\kappa_{1}(1-2\sqrt{\eta}\cos{\phi})+\kappa_{2}$ and $\kappa_{f,\text{eff}}=\kappa_{f}(1-2\sqrt{\eta}\cos{\phi})$ are the effective decay rates of the cavity and Fano modes; $G_{\text{eff}}=\lambda-i\kappa_{1f}[1-2\sqrt{\eta}\text{exp}(i\phi)]$ is the effective coupling coefficient between the cavity and Fano modes.
Moreover, the coherent feedback also modifies the noise terms in Eqs.~(\ref{aFanoSF}) and (\ref{fFanoSF}) but in a \textit{different} way from the decay terms. They are therefore not expressed as functions of the above-introduced effective decay rates.
Instead, the effective input noises upon the Fano mirror are given by input noise superpositions depending on the feedback efficiency and the phase accumulation. Interestingly, the coupling to the input signals is in general complex due to the coherent feedback.

In the presence of the single-sided feedback, the complex eigenvalues of the two optical normal modes are also strongly impacted. They become
\begin{equation}
\begin{split}
\tilde{\omega}_{\pm}&=\frac{\Delta_{a,\text{eff}}+\Delta_{f,\text{eff}}}{2}-i\frac{\kappa_{a,\text{eff}}+\kappa_{f,\text{eff}}}{2}\\
&\quad\,\pm\sqrt{\left(\frac{\Delta_{a,\text{eff}}-\Delta_{f,\text{eff}}}{2}-i\frac{\kappa_{a,\text{eff}}-\kappa_{f,\text{eff}}}{2}\right)^{2}+G_{\text{eff}}^{2}}\ .
\end{split}
\label{normalmodes}
\end{equation}
In this case, both the resonance frequencies [i.e., $\omega_{\pm}=\text{Re}(\tilde{\omega}_{\pm}$)] and the linewidths [i.e., $\kappa_{\pm}=-\text{Im}(\tilde{\omega}_{\pm}$)] of the two normal modes are modified by the coherent feedback.
Note that in Eq.~(\ref{normalmodes}) the optomechanical interactions between the mechanical and the two optical modes are not taken into account. However, the interference between the two optomechanical interaction paths may further enhance the Fano resonance~\cite{GenesQST}.

In view of this, Eq.~(\ref{normalmodes}) only provides an intuitive picture to understand the cooperation of the Fano resonance and the coherent feedback, rather than accurately predicting the optimal cooling region.
Nevertheless, we find that the optimal cooling region always appears when the real part of one of the normal-mode eigenenergies is comparable to the mechanical frequency, implying that the corresponding normal mode is driven close to the mechanical red sideband, and the linewidth of this normal mode is smaller than the mechanical frequency. 

Importantly, ground-state cooling of the mechanical mode would also be allowed by exploiting single-sided coherent feedback \textit{without} the Fano mode. However, the effective linewidth of the cavity mode that would be achieved by this is limited by the smallest of the two cavity decay rates (at the two end mirrors), thereby requiring that at least one of the mirrors has a decay rate that is smaller than the mechanical frequency. This is in contrast to the \textit{combined} outcome of coherent feedback and Fano mirror, where sideband cooling becomes possible even if all original decay rates of the cavity are far larger than the mechanical frequency, see Fig.~\ref{SL1}.

Before proceeding, we would like to briefly comment on another kind of coherent feedback that is mediated by only a few (or even a single) discrete modes.
At a first glance, the single-sided feedback loop can be simply realized by placing a vertical HRM on the same side of the Fano mirror, as shown in Fig.~\figpanel{STS}{a} in Appendix~\ref{appSW}.
However, we point out that this structure corresponds to a \textit{standing-wave} version of the feedback loop,
which could be viewed as a \textit{membrane-in-the-middle} optomechanical system if the HRM is very close to the Fano mirror, while it leads to \textit{non-Markovian} coherent feedback if the separation distance between the HRM and the Fano mirror is large enough. In contrast, a well-designed traveling-wave feedback loop as suggested here is preferable to implement instantaneous coherent feedback (for more details see Appendix~\ref{appSW}).

\subsection{Feedback-assisted ground-state cooling}\label{secSgb}

As in the previous section, one can determine the final phonon number from the quantum Langevin equations~\eqref{eq:FanoLangevin}
by solving the corresponding Lyapunov equation, see Appendix~\ref{appLyap}.
\begin{figure}[ptb]
\centering
\includegraphics[width=1.0\linewidth]{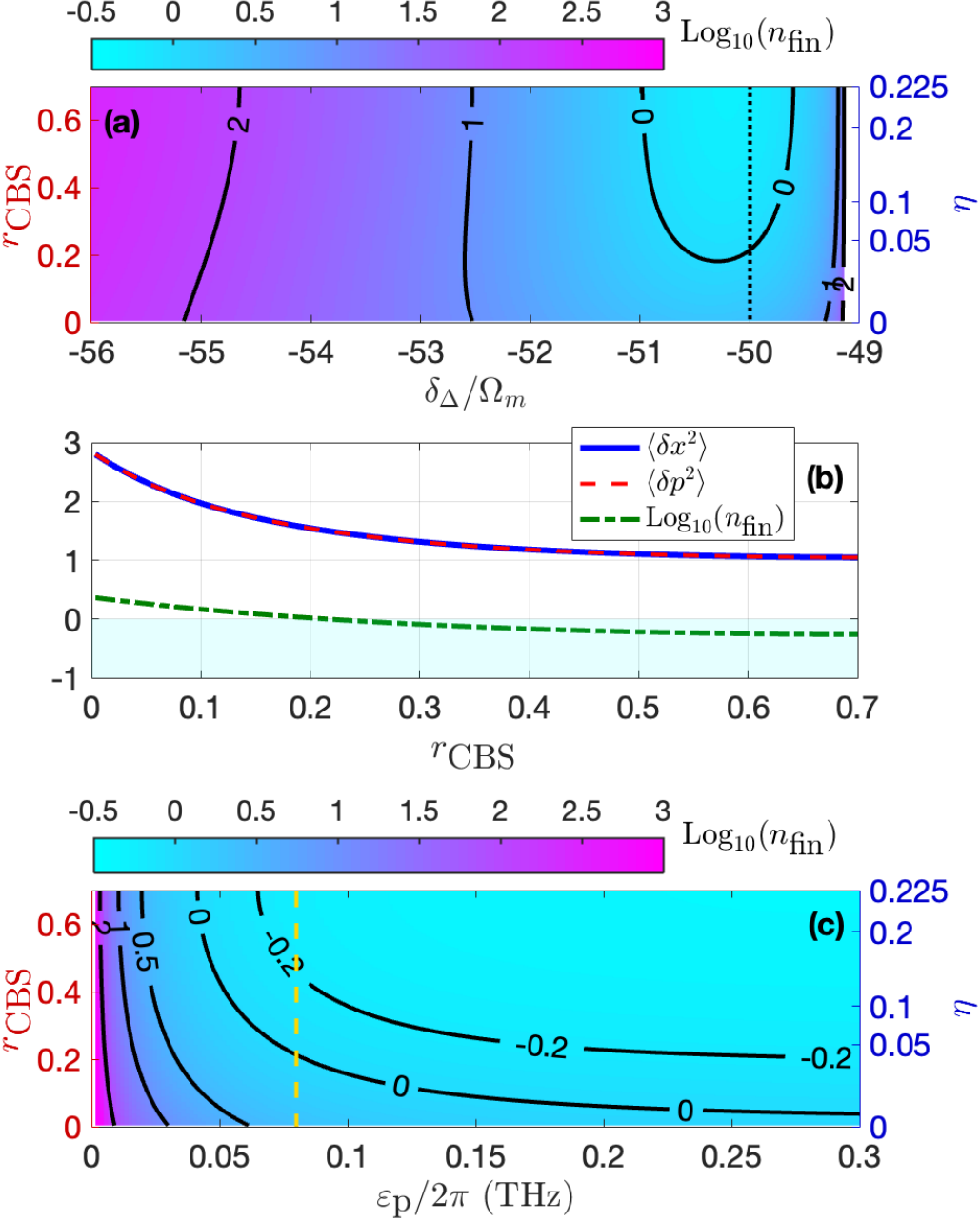}
\caption{Plots of the final phonon number $n_\mathrm{fin}$ (on logarithmic scale) for the single-sided feedback scheme. (a) Final phonon number versus reflection coefficient $r_\mathrm{CBS}$ (respectively the corresponding efficiency $\eta$) and detuning $\delta_{\Delta}$ with $\varepsilon_\mathrm{p}/2\pi=80\,\text{GHz}$. (b) Final phonon number and mechanical variances ($\langle\delta x^{2}\rangle$ and $\langle\delta p^{2}\rangle$) versus reflection coefficient $r_\mathrm{CBS}$ with $\varepsilon_\mathrm{p}/2\pi=80\,\text{GHz}$ and $\delta_{\Delta}/\Omega_{m}=-50$ [corresponding to the black dotted line in panel (a)]. The cyan area represents the regime of $n_\mathrm{fin}<1$. (c) Final phonon number versus reflection coefficient $r_\mathrm{CBS}$ and pumping amplitude $\varepsilon_\mathrm{p}$ with $\delta_{\Delta}/\Omega_{m}=-50$. The yellow dashed line corresponds to the driving amplitude used in (a) and (b). 
Other parameters are $\Omega_{m}/2\pi=1.3\,\text{MHz}$, $\kappa_{1}/2\pi=20\,\text{THz}$, $\kappa_{2}/2\pi=0.6\,\text{GHz}$, $\kappa_{f}/2\pi=1.08\,\text{GHz}$, $\gamma_{m}/2\pi=5\times 10^{-3}\,\text{Hz}$, $\lambda/2\pi=7\,\text{GHz}$, $\Delta_{a}/\Omega_{m}=30$, $g_{a,0}/\Omega_{m}=6.5\times 10^{-5}$, $g_{f,0}/\Omega_{m}=-1.6\times 10^{-4}$, $\phi=\pi$, $\eta_{\text{ex}}=0.9$, and $\bar{n}_{m}=10^{5}$.}
\label{SL1}
\end{figure}

Since the coherent feedback is introduced to the left Fano mirror in this case, we assume that the pumping field is applied to the right normal mirror. This allows the pumping field and the coherent feedback to be tuned independently, thus offering more flexibility for our cooling scheme. Specifically, this means that, in contrast to the previously shown two-sided feedback scheme, here the pumping amplitude $\varepsilon_\mathrm{p}$ is not modified by the feedback loop. The semiclassical steady-state values of the cavity and Fano modes are then
\begin{subequations}
\begin{eqnarray}
\bar{a}&=&\frac{\chi_{f,\text{eff}}^{-1}}{\chi_{a,\text{eff}}^{-1}\chi_{f,\text{eff}}^{-1}+G_{\text{eff}}^{2}}\varepsilon_\mathrm{p}, \label{alphas}\\
\bar{f}&=&\frac{-iG_{\text{eff}}}{\chi_{a,\text{eff}}^{-1}\chi_{f,\text{eff}}^{-1}+G_{\text{eff}}^{2}}\varepsilon_\mathrm{p}, \label{fss}
\end{eqnarray}
\end{subequations}
where
\begin{subequations}
\begin{eqnarray}
\chi_{a,\text{eff}}^{-1}&=&i\Delta_{a,\mathrm{eff}}+\kappa_{a,\mathrm{eff}}, \\
\chi_{f,\text{eff}}^{-1}&=&i\Delta_{f,\mathrm{eff}}+\kappa_{f,\mathrm{eff}}.
\end{eqnarray}
\end{subequations}
From Eqs.~\eqref{alphas} and \eqref{fss}, the enhanced optomechanical coupling coefficients can be obtained as $g_{a}=g_{a,0}\bar{a}$ and $g_{f}=g_{f,0}\bar{f}$. They are hence, due to the form of $\bar{a}$ and $\bar{f}$, generally not simultaneously real.

In Figs.~\figpanel{SL1}{a} and \figpanel{SL1}{b}, we show the dependence of the final phonon number $n_{\text{fin}}$ on the detuning $\delta_{\Delta}=\Delta_{a}-\Delta_{f}=\omega_{a}-\omega_{f}$ between the cavity and Fano modes and the feedback efficiency $\eta$ of the single-sided feedback loop, choosing a set of experimentally available parameters of such systems~\cite{Juliette2023arxiv} and a realistic initial phonon occupation $10^{5}$ (corresponding to a phonon bath of about $6\,\text{K}$), as indicated in the figure caption.
Note that in the ideal case of $\eta_{\text{ex}}=1$, the efficiency $\eta$ increases with $r_\mathrm{CBS}$ and reaches its maximum $\eta=0.25$ at $r_\mathrm{CBS}=\sqrt{0.5}\approx0.71$. However, here we consider a more realistic situation with $\eta_{\text{ex}}=0.9$, which accounts for a small but unavoidable loss in the feedback loop. With an appropriate feedback, ground-state cooling is allowed even in this deeply unresolved-sideband regime (the linewidth of the cavity resonance is more than seven orders of magnitude larger than the mechanical frequency). In contrast to the double-sided feedback scheme, here we assume $\phi=\pi$ (modulo $2\pi$) to facilitate ground-state cooling, the reason of which will be elucidated below.
The region of $n_\mathrm{fin}<1$, namely where ground-state cooling can be achieved, is reached when the real part of the normal-mode eigenenergy with the smallest imaginary part, here $\tilde{\omega}_{-}$, is comparable, but not exactly equal, to the mechanical frequency. Indeed, the minimum of $n_\mathrm{fin}$ is approached when $\omega_{-}/\Omega_{m}\approx1.1$, which is very close to the resonance condition. The slight deviation from the effective sideband cooling condition, $\omega_{-}=\Omega_{m}$, arises from the complicated interference effects of the optomechanical interactions with the coherent feedback. In Fig.~\figpanel{SL1}{b}, we also check that the equipartition of energy is approximately fulfilled within the region of $n_\mathrm{fin}<1$.

\begin{figure}[ptb]
\centering
\includegraphics[width=8.6 cm]{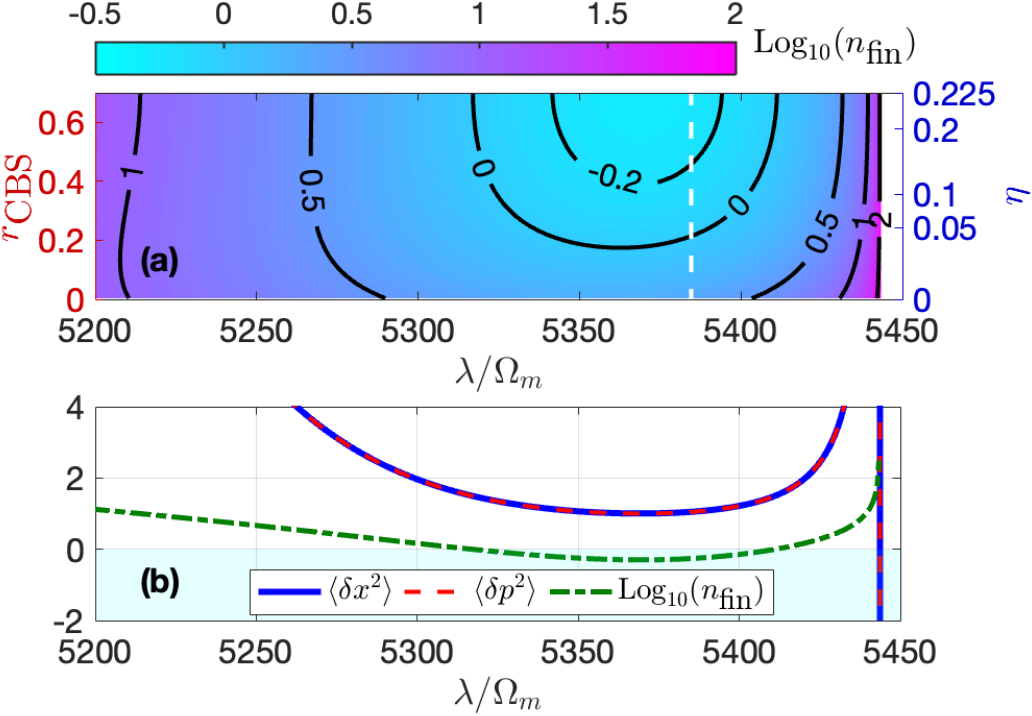}
\caption{(a) Final phonon number $n_\mathrm{fin}$ (on logarithmic scale) for the single-sided feedback scheme versus cavity-Fano coherent coupling strength $\lambda$ and reflection coefficient $r_\mathrm{CBS}$. The white dashed line correspond to the value of $\lambda$ used in Fig.~\ref{SL1}. (b)  Final phonon number $n_\mathrm{fin}$ (on logarithmic scale) and mechanical variances ($\langle\delta x^{2}\rangle$ and $\langle\delta p^{2}\rangle$) versus $\lambda$ with $r_{\textrm{CBS}}=0.7$. We assume $\delta_{\Delta}/\Omega_{m}=-50$ and other parameters are identical to those in Fig.~\figpanel{SL1}{a}.}
\label{SL2}
\end{figure}

As in a standard sideband cooling scheme, the cooling effect can be further improved by appropriately increasing the power of the pumping field, as shown in Fig.~\figpanel{SL1}{c}, yet one should be very careful since the system will enter the unstable region with strong enough pumping and large feedback efficiency. Moreover, we show in Fig.~\figpanel{Equip}{c} that the equipartition of energy can be significantly \emph{broken} with a strong pumping, even if the system is still in the stable regime. In view of this, the coherent feedback provides a way to enhance the cooling effect without increasing the pumping amplitude (i.e., without breaking the equipartition of energy).

In Fig.~\ref{SL1}, we have focused on the specific Fano-mirror setup where the photonic crystal membrane supports both the Fano and mechanical modes such that they are also coupled in a dispersive manner. We would like to point out that the cooling enhancement, which is based on the combination of the Fano resonance and the coherent feedback, can be extended to setups where the Fano and mechanical modes are decoupled~\cite{AE1,AE2,AE3,GenesDoped2014,FanoM,GenesQST,2cavity1,2cavity2,2cavity3,YCLiu2015pra}. A proof-of-principle demonstration of ground-state cooling in such setups can be found in Appendix~\ref{decouple}. We furthermore discuss in Appendix~\ref{appCCS} the role of the dissipative coupling $i\kappa_{1f}$ between the two optical modes, which is not present, e.g., in coupled-cavity cooling schemes \cite{2cavity1,2cavity2,2cavity3,YCLiu2015pra}.

In practice, there are many factors, e.g., imperfections of the fabrication of the cavity and photonic crystal that may affect the parameters that determine whether ground-state cooling becomes possible or not. In particular, the coherent coupling between the cavity and Fano modes, $\lambda$, depends on many factors, such as the specific material and structure of the Fano mirror as well as the cavity length.
This may pose challenges in precisely controlling the Fano resonance and thereby whether the mechanical mode can be cooled down as desired.
In other words, the \textit{actual} coherent coupling strength may deviate from the expected value such that the Fano resonance may become difficult to access. Nevertheless, we demonstrate in Fig.~\figpanel{SL2}{a} that the ground-state cooling is robust against a moderate coupling deviation with otherwise fixed parameters---the white dashed line corresponds to $\lambda=2\pi\times7\,\text{GHz}$ (i.e., $\lambda/\Omega_{m}=5384.6$) which is chosen in Fig.~\ref{SL1}.
We also examine in Fig.~\figpanel{SL2}{b} the equipartition of energy, which is always guaranteed within the working region of $n_{\text{fin}}<1$.

\begin{figure}[ptb]
\centering
\includegraphics[width=8.6 cm]{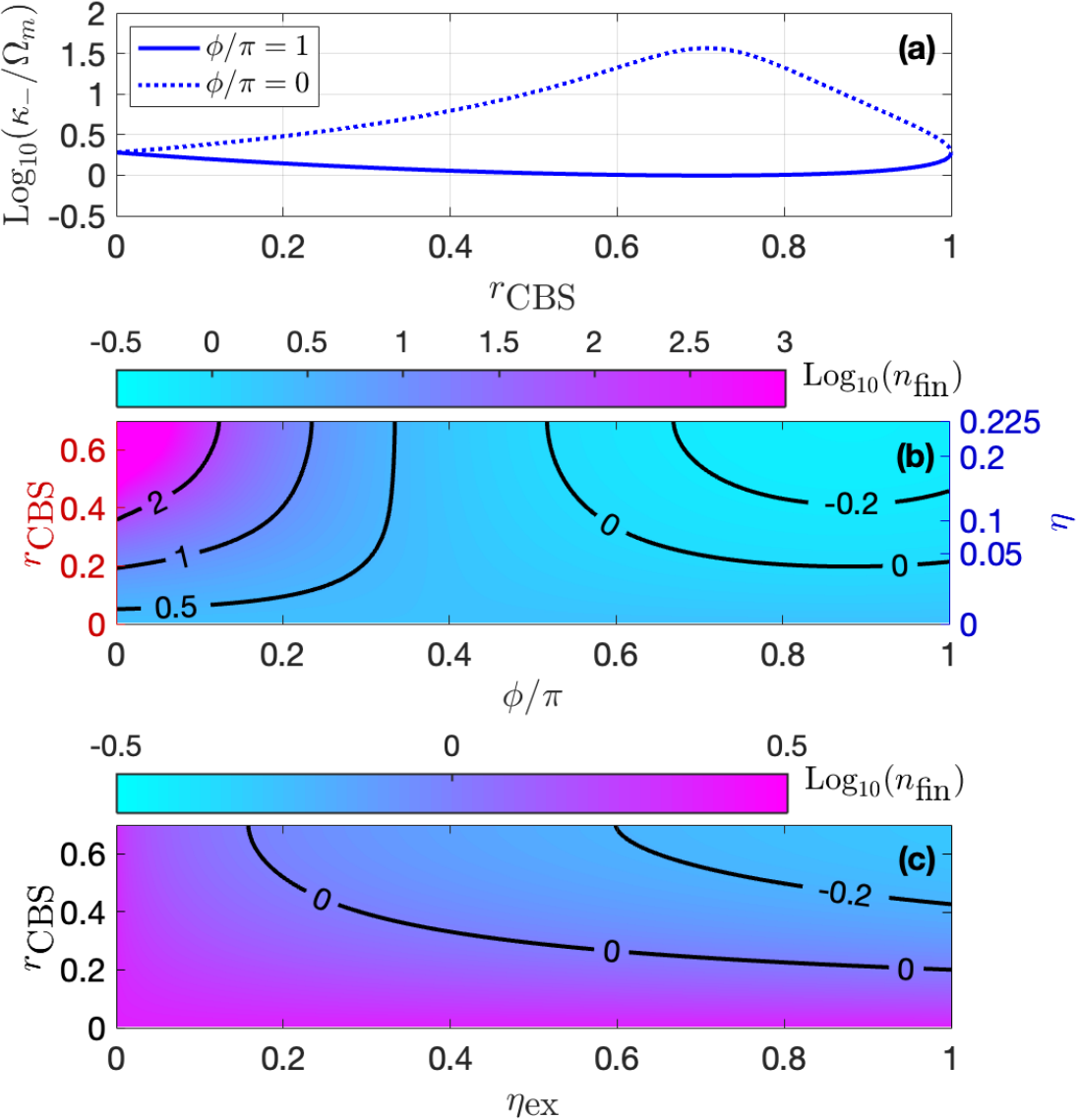}
\caption{(a) Linewidth $\kappa_{-}$ of the ``--'' normal mode as a function of reflection coefficient $r_{\text{CBS}}$ with $\phi/\pi=0$ and $\phi/\pi=1$. (b, c) Final phonon number $n_\mathrm{fin}$ (on logarithmic scale) versus reflection coefficient $r_\mathrm{CBS}$ and (b) phase accumulation $\phi$ and (c) extra efficiency factor $\eta_{\text{ex}}$. We assume $\eta_{\text{ex}}=0.9$ in (a) and (b) and $\phi=\pi$ in (c). Other parameters are identical to those in Fig.~\figpanel{SL1}{a}.}
\label{reason}
\end{figure}

As mentioned above, the cooling effect in the Fano-mirror setup can be enhanced by the single-sided feedback scheme when $\phi=\pi$. This is quite different from the double-sided feedback scheme, where the standard optomechanical system (without the Fano mode) enters the effective sideband-resolved regime when $\phi=0$. This difference arises from the \emph{interplay} between the Fano resonance and the coherent feedback. The former becomes highly effective with strong optical dissipative coupling $\kappa_{1f}$, which leads to a significant linewidth splitting between the two normal modes (as also discussed in Appendix~\ref{appCCS}). Importantly, the optical dissipative coupling would be significantly suppressed when $\phi=0$, due to the destructive interference arising from the coherent feedback, even if the feedback loop is not perfect ($\eta_{\text{ex}}<1$). This would cause the smaller normal-mode linewidth to increase, as shown in Fig.~\figpanel{reason}{a}. In contrast, by employing coherent feedback with constructive interference (e.g., $\phi=\pi$), the smaller normal-mode linewidth can be further reduced due to the effectively enhanced optical dissipative coupling. 

Finally, we would like to point out that the enhanced cooling effect can still be achieved when the phase accumulation $\phi$ deviates moderately from odd multiples of $\pi$, as shown in Fig.~\figpanel{reason}{b}. This demonstrates the robustness of our scheme against the practically unavoidable inaccuracy of the feedback loop length. Moreover, Fig.~\figpanel{reason}{c} shows that the single-sided feedback scheme remains effective even with a much stronger optical loss in the feedback loop. While we have always assumed $\eta_{\text{ex}}=0.9$ in the previous figures, it is evident that the coherent feedback can still facilitate ground-state cooling even when $\eta_{\text{ex}}=0.2$. 

These results show that the combination of a Fano mirror with single-sided coherent feedback is a promising scheme to achieve ground-state cooling circumventing too strict requirements on the parameters of the Fano-mirror optomechanical setup. However, further optimization of parameters can lead to even lower phonon numbers, e.g., when increasing the pumping amplitude $\varepsilon_{\text{p}}$, reducing the optical coherent coupling $\lambda$, or increasing the cavity decay $\kappa_{1}$ through the Fano mirror.

\section{Discussion and Conclusions}\label{secConc}
We have explored ground-state cooling of the mechanical mode of a Fano-mirror optomechanical system, which is allowed based on the \textit{cooperation} of an external coherent feedback with the Fano resonance (arising from the interaction between the cavity mode and the guided optical mode in the photonic crystal membrane).
While the interaction between the cavity and Fano modes can lead to an optical normal mode with a linewidth that is smaller than the mechanical frequency, it is in many experimentally relevant situations only thanks to the coherent feedback that ground-state cooling is enabled. Indeed, we have shown that within a broad parameter regime, coherent feedback further modifies the optical properties and input noises of the setup in order to reach the goal.

More specifically, we have shown that single-sided feedback---in contrast to the more standard two-sided feedback---is a good candidate for optomechanical setups featuring a two-sided cavity with large and \textit{very different} decay rates at the two cavity mirrors. As a result, the mechanical mode can be cooled down towards its ground state, even if the optomechanical system works in a deeply sideband-unresolved regime---with the total optical linewidth more than seven orders of magnitude larger than the mechanical frequency---and even if the feedback efficiency is low. An advantage for future experimental realizations, where precise parameter values might be hard to control, is also that, with the help of coherent feedback, ground-state cooling is fairly robust against some modifications in the experimental parameters with respect to their ideal value.

Realizations of ground-state cooling in conventional optomechanical setups, namely without a Fano resonance, are limited, both when using double- and single-sided coherent feedback schemes. Concretely, the double-sided coherent feedback becomes inapplicable if the cavity has very different decay rates at the two mirrors, while the single-sided coherent feedback is not sufficient to realize ground-state cooling on its own if the smallest original decay rate is larger than the mechanical frequency.

Here, we have shown that realizing ground-state cooling of the mechanical mode seems within experimental reach, in Fano-mirror optomechanical setups when only combining it with a low-efficiency single-sided coherent feedback. This is especially promising for microcavities with a photonic crystal mirror exhibiting very large single-photon optomechanical couplings~\cite{WWoe2023} since it could help access the nonlinear regime and pave the way for quantum technological applications such as quantum sensing.

\acknowledgments
We thank Vitaly Shumeiko, Yong Li, Anton Frisk Kockum, Anastasiia Ciers, Alexander Jung, and Hannes Pfeifer for helpful discussions. We acknowledge financial support by the Knut och Alice Wallenberg stiftelse through project grant no. 2022.0090, as well as through individual Wallenberg Academy fellowship grants (J.S., W.W.), and the QuantERA project C’MON-QSENS!.

\appendix

\counterwithin*{figure}{part}
\stepcounter{part}
\renewcommand{\thefigure}{\Alph{part}.\arabic{figure}}

\section{Implementations of the double-sided and single-sided feedback loops}\label{appSL}

\begin{figure*}[ptb]
\centering
\includegraphics[width=13 cm]{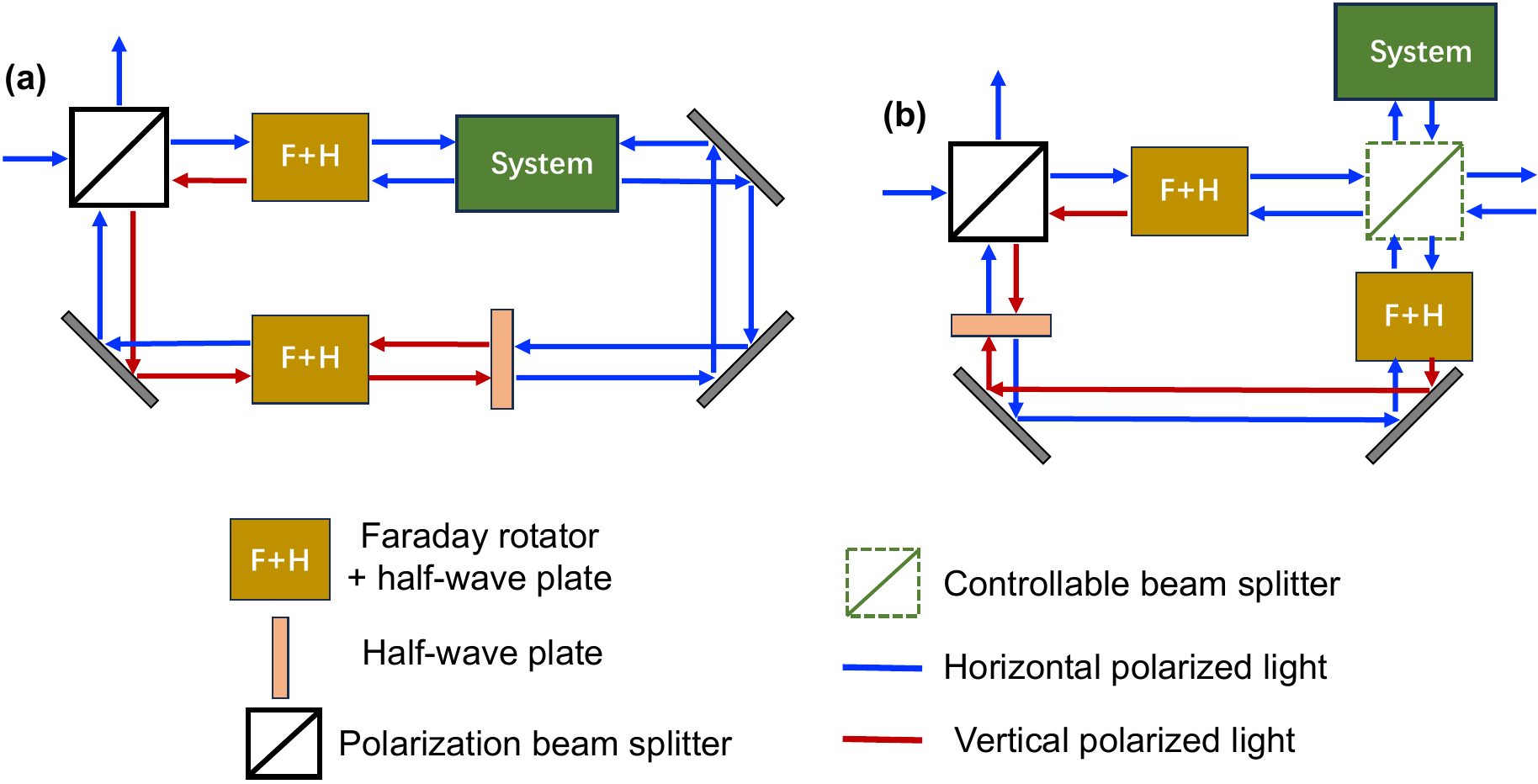}
\caption{Experimental implementations of (a) the double-sided and (b) the single-sided coherent feedback schemes.}
\label{Implementation}
\end{figure*}

In this appendix, we briefly discuss the possible experimental implementations of the double-sided and single-sided coherent feedback schemes considered in the main text. As shown in Figs.~\ref{Dbloop} and \ref{Sloop}, both schemes require optical circulators to couple the Fabry-Perot-type optomechanical system to a traveling-wave field twice. One possible implementation of such circulators is the combination of a polarization beam splitter, a Faraday rotator, and a well aligned half-wave plate, as shown in Figs.~\figpanel{Implementation}{a} and \figpanel{Implementation}{b}.

Note that the polarization change of light passing through a half-wave plate, typically consisting of a birefringent crystal, depends on the angle of its polarization plane with respect to the extraordinary axis, while the polarization change of a Faraday rotator is only determined by the magnetic field direction and the sign of the Verdet constant~\cite{Verdet2019}, describing the strength of the Faraday effect for a particular material. In view of this, with a well-designed combination of a Faraday rotator and a half-wave plate, the polarization of the light can remain unaffected in one direction but will change by $90^{\circ}$ in the opposite direction.

While the single-sided feedback scheme in Fig.~\figpanel{Implementation}{b} is lossy and thus the feedback efficiency is at most $0.25$, the double-sided feedback scheme in Fig.~\figpanel{Implementation}{a} shows a nearly $100\%$ efficiency, except for the unavoidable weak losses via the scattering and absorption processes.

\section{Methods for calculating the final phonon number}\label{appLyap}
\stepcounter{part}
\subsection{Double-sided feedback}

For an optomechanical system with linearized dynamics, the steady-state values of the correlations can be calculated by formulating a Lyapunov equation for the covariance matrix $V_{ij}(t)=\langle R_{i}(t)R_{j}(t)+R_{j}(t)R_{i}(t)\rangle/2$. This involves the quadrature vector $R=(\delta X_{a}, \delta P_{a}, \delta x, \delta p)^{T}$ with $X_{a}=(a^{\dag}+a)/\sqrt{2}$ and $P_{a}=i(a^{\dag}-a)/\sqrt{2}$ being two orthogonal quadratures of the cavity mode.

More specifically, for the double-sided feedback scheme considered in Fig.~\ref{Dbloop}, the linearized quantum Langevin equations~(\ref{xFano})--(\ref{pFano}) and~(\ref{Dba2}) can be recast into the compact matrix form
\begin{equation}
    \frac{d}{dt}R=MR+U,
    \label{compactDb}
\end{equation}
with the $4\times4$ coefficient matrix, also called drift matrix,
\begin{eqnarray*}
M=
-{\left(\begin{array}{cccccc}
\kappa_{\text{eff}} & -\Delta_{\text{eff}} & 0 & 0 \\
\Delta_{\text{eff}} & \kappa_{\text{eff}} & \sqrt{2}g_{a} & 0 \\
0 & 0 & 0 & -\Omega_{m} \\
\sqrt{2}g_{a} & 0 & \Omega_{m} & \gamma_{m}
\end{array}\right)},
\label{Mmatrix2}
\end{eqnarray*}
and the vector of noise quadratures
\begin{widetext}
\begin{eqnarray*}
U=
\left(\begin{array}{c}
(\sqrt{2\kappa_{1}}-\sqrt{2\kappa_{2}\eta}\cos{\phi})X_{1}^{\text{in}}+\sqrt{2\kappa_{2}\eta}\sin{\phi}P_{1}^{\text{in}}+\sqrt{2\kappa_{2}(1-\eta)}X_{2'}^{\text{in}} \\
(\sqrt{2\kappa_{1}}-\sqrt{2\kappa_{2}\eta}\cos{\phi})P_{1}^{\text{in}}-\sqrt{2\kappa_{2}\eta}\sin{\phi}X_{1}^{\text{in}}+\sqrt{2\kappa_{2}(1-\eta)}P_{2'}^{\text{in}} \\
0 \\
\sqrt{2\gamma_{m}}\xi_{m}
\end{array}\right)\ .
\end{eqnarray*}
\end{widetext}
Note that here we have assumed a real $\bar{a}$ without loss of generality, which can always be achieved by tuning the phase of the pumping field.
Then, the covariance matrix $V$ obeys the evolution equation $dV/dt=MV+VM^{T}+N$, where
\begin{eqnarray}
N=\left(\begin{array}{cccccc}
\kappa_{\text{tot},\text{eff}} & 0 & 0 & 0 \\
0 & \kappa_{\text{tot},\text{eff}} & 0 & 0 \\
0 & 0 & 0 & 0 \\
0 & 0 & 0 & \gamma_{m}(2n_{m}+1)
\end{array}\right)
\label{Nmatrix}
\end{eqnarray}
is the diffusion matrix satisfying $N_{ij}\delta(t-t')=\langle U_{i}(t)U_{j}(t')+U_{j}(t')U_{i}(t)\rangle/2$.
In our linearized optomechanical system, the final phonon number,
\begin{equation}
    n_\mathrm{fin}=\frac{1}{2}\left(V_{33}+V_{44}-1\right),
\end{equation}
can be numerically determined by solving the steady-state Lyapunov equation,
\begin{equation}
    MV+VM^{T}=-N.
    \label{LyapE}
\end{equation}

\subsection{Single-sided feedback}

For the single-sided feedback scheme considered in Fig.~\ref{Sloop}, the quantum Langevin equations, Eqs.~(\ref{xFano})--(\ref{pFano}) and Eqs.~(\ref{aFanoSF})--(\ref{fFanoSF}), can be recast into a similar form as that in Eq.~(\ref{compactDb}) for the quadrature vector $R'=(\delta X_{a}, \delta P_{a}, \delta X_{f}, \delta P_{f}, \delta x, \delta p)^{T}$,  with a $6\times6$ drift matrix
\begin{widetext}
\begin{eqnarray}
M'=-{\left(\begin{array}{cccccc}
\kappa_{a,\text{eff}} & -\Delta_{a,\text{eff}} & -\text{Im}(G_{\text{eff}}) & -\text{Re}(G_{\text{eff}}) & -\sqrt{2}\text{Im}(g_{a}) & 0 \\
\Delta_{a,\text{eff}} & \kappa_{a,\text{eff}} & \text{Re}(G_{\text{eff}}) & -\text{Im}(G_{\text{eff}}) & \sqrt{2}\text{Re}(g_{a}) & 0 \\
-\text{Im}(G_{\text{eff}}) & -\text{Re}(G_{\text{eff}}) & \kappa_{f,\text{eff}} & -\Delta_{f,\text{eff}} & -\sqrt{2}\text{Im}(g_{f}) & 0 \\
\text{Re}(G_{\text{eff}}) & -\text{Im}(G_{\text{eff}}) & \Delta_{f,\text{eff}} & \kappa_{f,\text{eff}} & \sqrt{2}\text{Re}(g_{f}) & 0 \\
0 & 0 & 0 & 0 & 0 & -\Omega_{m} \\
\sqrt{2}\text{Re}(g_{a}) & \sqrt{2}\text{Im}(g_{a}) & \sqrt{2}\text{Re}(g_{f}) & \sqrt{2}\text{Im}(g_{f}) & \Omega_{m} & \gamma_{m}
\end{array}\right)}
\label{M'matrix2}
\end{eqnarray}
and a corresponding noise vector
\begin{eqnarray}
U'=
\left(\begin{array}{c}
\sqrt{2\kappa_{1}}W_{1}X_\mathrm{top}^{\text{in}}+\sqrt{2\kappa_{1}}W_{2}P_\mathrm{top}^{\text{in}}+\sqrt{2\kappa_{1}}W_{3}X_\mathrm{bot}^{\text{in}}-\sqrt{2\kappa_{1}}W_{4}P_\mathrm{bot}^{\text{in}}+\sqrt{2\kappa_{2}}X_{2}^{\text{in}} \\
\sqrt{2\kappa_{1}}W_{1}P_\mathrm{top}^{\text{in}}-\sqrt{2\kappa_{1}}W_{2}X_\mathrm{top}^{\text{in}}+\sqrt{2\kappa_{1}}W_{3}P_\mathrm{bot}^{\text{in}}+\sqrt{2\kappa_{1}}W_{4}X_\mathrm{bot}^{\text{in}}+\sqrt{2\kappa_{2}}P_{2}^{\text{in}} \\
\sqrt{2\kappa_{f}}W_{1}X_\mathrm{top}^{\text{in}}+\sqrt{2\kappa_{f}}W_{2}P_\mathrm{top}^{\text{in}}+\sqrt{2\kappa_{f}}W_{3}\tilde{X}_\mathrm{bot}^{\text{in}}-\sqrt{2\kappa_{f}}W_{4}P_\mathrm{bot}^{\text{in}} \\
\sqrt{2\kappa_{f}}W_{1}P_\mathrm{top}^{\text{in}}-\sqrt{2\kappa_{f}}W_{2}X_\mathrm{top}^{\text{in}}+\sqrt{2\kappa_{f}}W_{3}P_\mathrm{bot}^{\text{in}}+\sqrt{2\kappa_{f}}W_{4}X_\mathrm{bot}^{\text{in}} \\
0 \\
\sqrt{2\gamma_{m}}\xi_{m}
\end{array}\right).
\label{Svector}
\end{eqnarray}
\end{widetext}
In Eq.~(\ref{Svector}), we have defined the shorthand notations
$W_{1}=r_\mathrm{CBS}\left[1-\sqrt{\eta}\cos{(\phi)}\right]$, $W_{2}=r_\mathrm{CBS}\sqrt{\eta}\sin{(\phi)}$,
$W_{3}=t_\mathrm{CBS}\left[1-\sqrt{\eta}\cos{(\phi)}\right]$, $W_{4}=t_\mathrm{CBS}\sqrt{\eta}\sin{(\phi)}$.
In this case, the Lyapunov equation is given by
\begin{equation}
M'V'+V'M'^{\text{T}}+N'=0,
\label{SLyap}
\end{equation}
where
\begin{eqnarray}
N'=\left(\begin{array}{cccccc}
F_{1}' & 0 & F_{2}' & 0 & 0 & 0 \\
0 & F_{1}' & 0 & F_{2}' & 0 & 0 \\
F_{2}' & 0 & F_{3}' & 0 & 0 & 0 \\
0 & F_{2}' & 0 & F_{3}' & 0 & 0 \\
0 & 0 & 0 & 0 & 0 & 0 \\
0 & 0 & 0 & 0 & 0 & \gamma_{m}(2n_{m}+1)
\end{array}\right)
\label{Nmatrix2}
\end{eqnarray}
with
\begin{eqnarray*}
F_{1}'&=&\left(W_{1}^{2}+W_{2}^{2}+W_{3}^{2}+W_{4}^{2}\right)\kappa_{1}+\kappa_{2}, \label{f1p}\\
F_{2}'&=&\left(W_{1}^{2}+W_{2}^{2}+W_{3}^{2}+W_{4}^{2}\right)\kappa_{1f}, \label{f2p} \\
F_{3}'&=&\left(W_{1}^{2}+W_{2}^{2}+W_{3}^{2}+W_{4}^{2}\right)\kappa_{f}. \label{f3p}
\end{eqnarray*}
Then the final phonon number of the mechanical mode can be determined as
\begin{eqnarray}
n_\mathrm{fin}&=&\frac{1}{2}(\langle\delta x^{2}\rangle+\langle\delta p^{2}\rangle-1) \nonumber\\
&=&\frac{1}{2}\left(V'_{55}+V'_{66}-1\right).
\label{V5566}
\end{eqnarray}

\section{Equipartition of energy in the ground-state cooling regimes}\label{appEqp}
\stepcounter{part}
As mentioned in the main text, another important criterion of the mechanical ground-state cooling, besides $n_\mathrm{fin}<1$, is the equipartition of energy, i.e., $\langle\delta x^{2}\rangle\simeq\langle\delta p^{2}\rangle$.
Otherwise the steady state of the system is not a strict thermal equilibrium state and thus there is not a well-defined effective temperature of the mechanical mode in this case~\cite{Genes2008pra}. Therefore, it is also necessary to examine the two variances $\langle\delta x^{2}\rangle$ and $\langle\delta p^{2}\rangle$ of the mechanical mode in the regimes of $n_\mathrm{fin}<1$ and and see if a genuine ground-state cooling can be expected.

\begin{figure}[ptb]
\centering
\includegraphics[width=8.5 cm]{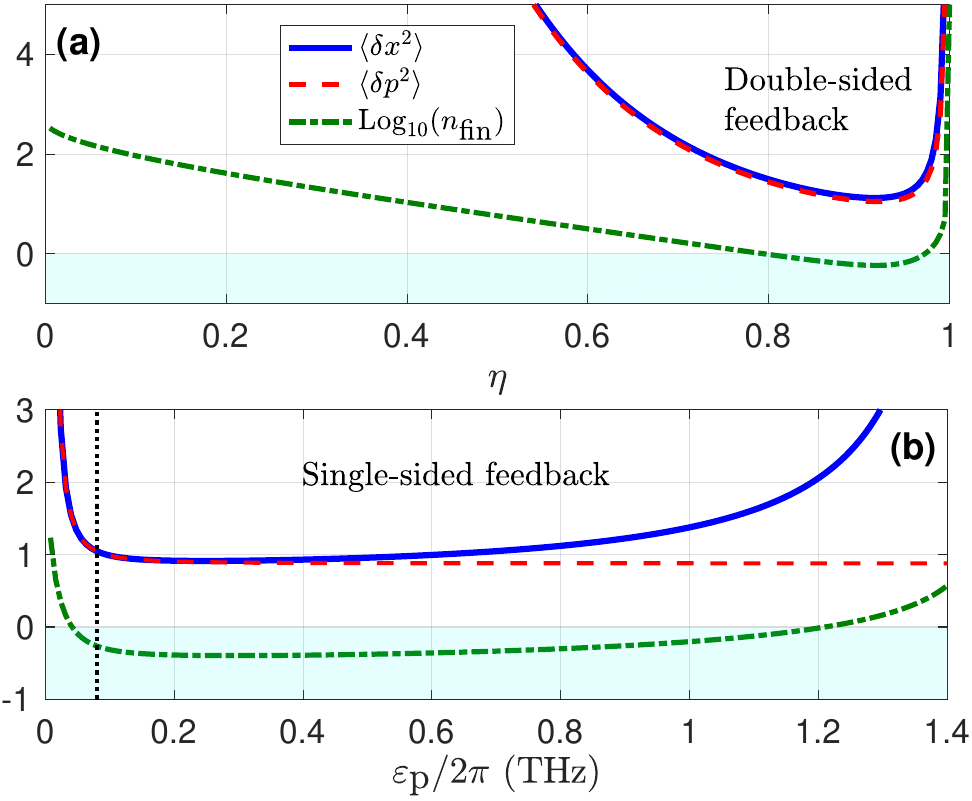}
\caption{(a) Final phonon number $n_\mathrm{fin}$ (on logarithmic scale) and mechanical variances ($\langle\delta x^{2}\rangle$ and $\langle\delta p^{2}\rangle$) versus feedback efficiency $\eta$ for the double-sided feedback scheme with $\phi=0$. (b) Final phonon number and mechanical variances versus pumping amplitude $\varepsilon_{\text{p}}$ for the single-sided feedback scheme with $r_{\text{CBS}}=0.7$. The two panels share the same legend. The cyan area represents the regime of $n_\mathrm{fin}<1$ and the black dotted line in (b) indicates the pumping amplitude used in Fig.~\figpanel{SL1}{a}. Other parameters in (a) and (b) are identical with those in Fig.~\figpanel{DbNf}{b} and Fig.~\figpanel{SL1}{c}, respectively.}
\label{Equip}
\end{figure}

We first present an example for the double-sided feedback scheme in Fig.~\figpanel{Equip}{a}, which corresponds to the orange solid line in Fig.~\figpanel{DbNf}{b}. We find that the two mechanical variances are in good agreement, with their difference being negligible compared to their average values. For the single-sided feedback scheme, we show in Fig.~\figpanel{Equip}{c} that the difference between the two mechanical variances increases gradually with $\varepsilon_{\text{p}}$. This indicates that increasing the pumping amplitude is not always helpful for enhancing the cooling effect, even if it results in a lower final phonon number, as is shown in Fig.~\figpanel{SL1}{c}. In view of this, we choose an appropriate value for $\varepsilon_{\text{p}}$ in all other panels of Figs.~\ref{SL1}--\ref{reason}, as indicated by the black dotted line in Fig.~\figpanel{Equip}{b}, to ensure the equipartition of energy.

\section{Optical normal modes with double-sided coherent feedback}\label{appNormal}
\stepcounter{part}

\begin{figure}[pt]
\centering
\includegraphics[width=0.9\linewidth]{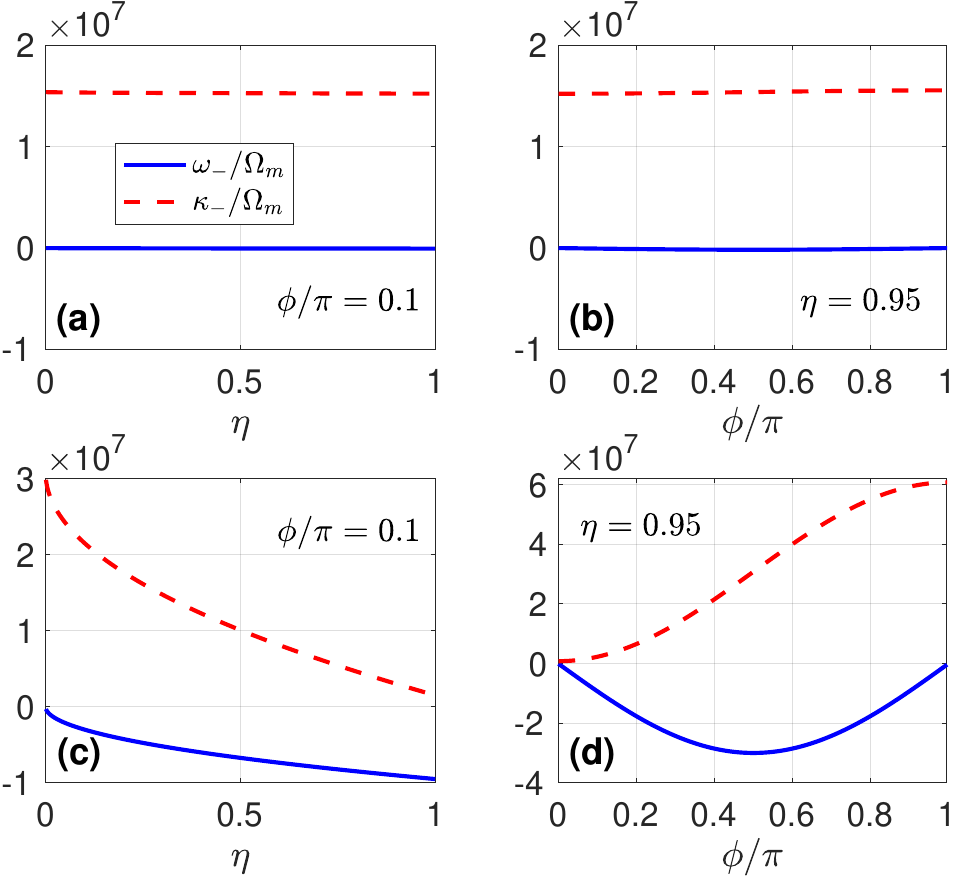}
\caption{Effective resonance frequency $\omega_{-}$ and linewidth $\kappa_{-}$ of the ``--'' optical normal mode versus the feedback efficiency $\eta$ and phase accumulation $\phi$ for $\kappa_{2}/2\pi=0.6\,\text{GHz}$ [(a) and (b)] and $\kappa_{2}/2\pi=20\,\text{THz}$. All the panels share the same legend. Other parameters, except for those indicated in the panels, are $\kappa_{1}/2\pi=20\,\text{THz}$, $\kappa_{f}/2\pi=1.08\,\text{GHz}$, $\Delta_{a}/\Omega_{m}=30$, $\delta_{\Delta}/\Omega_{m}=18.2$, $\lambda/2\pi=805\,\text{MHz}$.}\label{DWFano}
\end{figure}

In this appendix, we aim to show that the double-sided coherent feedback has a negligible impact on the Fano-mirror optomechanical setup when the cavity is highly asymmetric (i.e., $\kappa_{2}\ll\kappa_{1}$). This conclusion can be drawn from examining the two optical normal modes, in this case formed by the cavity and Fano modes. 

When including the Fano mode in the double-sided-feedback scheme, the quantum Langevin equations of the two optical modes are given by
\begin{eqnarray*}
\delta\dot{a}&=&-(i\Delta_{a,\text{eff}}+\kappa_{\text{tot}, \text{eff}})\delta a-iG'\delta f-ig_{a}\delta x \\
&&+\left(\sqrt{2\kappa_{1}}-\sqrt{2\kappa_{2}\eta}e^{i\phi}\right)a_{1}^{\text{in}}+\sqrt{2\kappa_{2}(1-\eta)}a_{2}^{\text{in}'}, \\
\delta\dot{f}&=&-(i\Delta_{f}+\kappa_{f})\delta f-iG\delta a -ig_{f}\delta x+\sqrt{2\kappa_{f}}a_{1}^{\text{in}},
\end{eqnarray*}
where $G'=G+2i\kappa_{2f}\sqrt{\eta}e^{i\phi}$ and all other symbols are defined as in the main text. Note that the overall interaction between the cavity and Fano modes becomes asymmetric since the unidirectional traveling-wave feedback loop leads to a ``cascaded'' interaction. In this case, the eigenfrequencies of the two normal modes are obtained as
\begin{equation}
\begin{split}
\tilde{\omega}_{\pm}&=\frac{\Delta_{a,\text{eff}}+\Delta_{f}}{2}-i\frac{\kappa_{\text{tot},\text{eff}}+\kappa_{f}}{2}\\
&\quad\,\pm\sqrt{\left(\frac{\Delta_{a,\text{eff}}-\Delta_{f}}{2}-i\frac{\kappa_{\text{tot},\text{eff}}-\kappa_{f}}{2}\right)^{2}+GG'}.
\end{split}
\end{equation}

As an example, Figs.~\figpanel{DWFano}{a} and \figpanel{DWFano}{b} [Figs.~\figpanel{DWFano}{c} and \figpanel{DWFano}{d}] show that the effective resonance frequency $\omega_{-}$ and linewidth $\kappa_{-}$ of the ``--'' normal mode are barely affected (significantly affected) by the coherent feedback when $\kappa_{2}\ll\kappa_{1}$ (when $\kappa_{2}=\kappa_{1}$). One can thus conclude that in the highly asymmetric case of $\kappa_{2}\ll\kappa_{1}$, the double-sided coherent feedback is inefficient for the Fano-mirror optomechanical setup.
\color{black}

\section{Standing-wave and traveling-wave versions of the single-sided feedback loop}\label{appSW}
\stepcounter{part}

A single-sided feedback loop can, as
mentioned in the main text, be realized in very different ways, depending on the specific configuration (standing-wave or traveling-wave versions) as well as on the length of the feedback loop. In this appendix, we first provide general descriptions for the interaction between the optomechanical system and the ``feedback loop'' (i.e., the modes in the loop that are coupled to the cavity; hereafter we refer to them as ``loop modes''), and then identify the conditions under which various theoretical descriptions are applicable. We here treat a simple system with a single optical mode $a$; generalizations to multiple optical modes, e.g., additional Fano modes, are straightforward.

\subsection{Standing-wave version}
We first consider a \textit{standing-wave} version of the single-sided coherent feedback, in contrast to the traveling-wave version considered in the main text and in Appendix~\ref{sec_traveling}. As shown in Fig.~\figpanel{STS}{a}, the feedback loop is formed by simply placing a vertical HRM on one side of the cavity (here it is placed on the same side of the Fano mirror) at a distance $d$.
Such a model is equivalent to a \textit{direct-coupled structure}~\cite{directcouple} in wave-guide quantum electrodynamics, where the field is terminated at the boundary of the system. We assume that the optomechanical system is coupled to $j_{\text{max}}$ loop modes in total. The Hamiltonian of the whole model (system plus loop) can be written as $H_{\text{tot}}=H_{\text{sys}}+H_{\text{loop}}+H_{\text{int}}$, where $H_{\text{sys}}$ is the Hamiltonian of the (Fano-mirror) optomechanical system,
\begin{equation}
H_{\text{loop}}=\sum_{j=1}^{j_{\text{max}}}\omega_{j}b_{j}^{\dag}b_{j}
\label{Hloop}
\end{equation}
is the Hamiltonian of the loop modes, and
\begin{equation}
\begin{split}
H_{\text{int}}&=\left[\sum_{j=1}^{j_{\text{max}}}\xi_{j}a^{\dag}\sin{(k_{j}d)}b_{j}+\text{H.c.}\right]+\sum_{j=1}^{j_{\text{max}}}H_{\text{om},j}
\end{split}
\label{HSL}
\end{equation}
describes the system-loop interaction. Here, $\omega_{j}$ ($k_{j}$) is the frequency (wave vector) of the $j$th loop mode $b_{j}$; $\xi_{j}$ is the coupling amplitude of loop mode $b_{j}$ and the cavity mode of the optomechanical system; the sinusoidal function $\sin{(k_{j}d)}$ in Eq.~(\ref{HSL}) results from the wave functions of the \textit{standing modes} in the loop space (we assume that the HRM is placed at $x=0$ without loss of generality), which is reminiscent of the case of an atom in front of a mirror~\cite{Xuereb2009,FCmirror}; $H_{\text{om},j}$ is the term describing the optomechanical coupling between $b_{j}$ and the mechanical mode, with the specific form depending on many factors (the distance $d$, the reflectivity and position of the mirror etc.) as will be discussed below.

\subsubsection{Markovian reservoir limit}

In the limit of $d\rightarrow+\infty$, the free spectral range $\omega_{\text{FSR}}=\pi c/d$ of the loop modes approaches zero such that the cavity and Fano modes of the optomechanical system are coupled to \textit{a continuum of modes}. In this case, the feedback time goes to infinity, $\tau=2d/c\rightarrow+\infty$, implying that the system is coupled to a \textit{Markovian} reservoir with no coherent feedback. Moreover, in this case one can just assume $H_{\text{om},j}\rightarrow0$ since the mechanical oscillation of the Fano mirror has a negligible influence on the reservior.
Now the Hamiltonian~(\ref{HSL}) becomes
\begin{equation}
H_{\text{int,Markov}}=\sum_{j=1}^{+\infty}\xi_{j}a^{\dag}b_{j}+\text{H.c.},
\label{HSLmk}
\end{equation}
where we have assumed constant system-reservoir coupling amplitudes based on the Weisskopf-Wigner approximation. Moreover, in Eq.~(\ref{HSLmk}) we have removed the sinusoidal-function dependence of the coupling amplitudes, since the loop modes are clearly not standing-wave modes in this case.

\begin{figure}[ptb]
\centering
\includegraphics[width=8.0 cm]{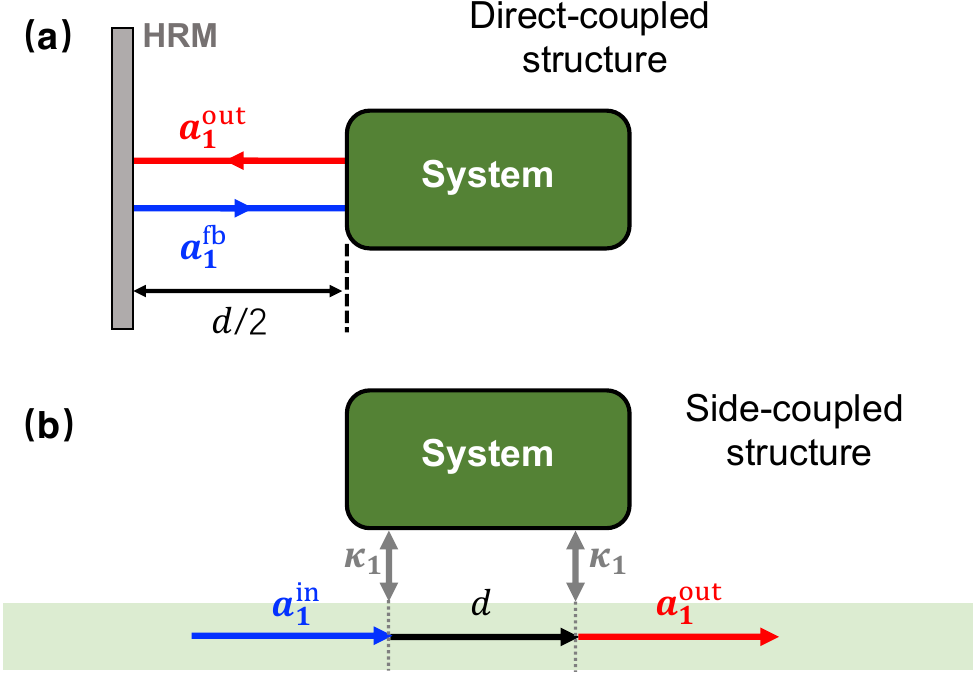}
\caption{(a) A standing-wave version of the single-sided coherent feedback, which is equivalent to a direct-coupled structure in waveguide quantum electrodynamics. (b) The equivalent side-coupled structure of the traveling-wave single-sided coherent feedback in Fig.~\ref{Sloop}.}\label{STS}
\end{figure}

\subsubsection{Membrane-in-the-middle setup limit}

When $d$ is very small, namely $d\sim L$ with $L$ the length of the optomechanical cavity, the hierarchy $\omega_{\text{FSR}}\gg\xi_{j}$ ensures that in the energy-window given by the linewidth of the cavity mode, there is only \textit{a single loop mode} (i.e., $j_{\text{max}}=1$). In this case, the cavity mode is hence coupled only to one loop mode and the whole setup is equivalent to a \textit{membrane-in-the-middle} optomechanical system~\cite{Harris2008nature}. Here, we consider the case, where the resonance condition $d=m\pi/k_b$ [i.e., $\sin{(k_b d)}\equiv 0$], with $k_b$ the wave number of the loop mode and $m$ an arbitrary positive integer,  is fulfilled. Then, another cavity is created on the left side of the mirror. Due to this choice the photonic tunneling terms (i.e., interactions between $b$ and $a$) in Eq.~(\ref{HSLmim}) disappear. As a consequence,  Eqs.~(\ref{Hloop}) and (\ref{HSL}) become
\begin{eqnarray}
H_{\text{loop,MIM}}&=&\omega_{b}b^{\dag}b, \label{Hloopmim} \\
H_{\text{int,MIM}}&=&H_{\text{om}}, \label{HSLmim}
\end{eqnarray}
where $\omega_{b}= ck_b$ represents the resonance frequency  of mode $b$.

For such a system, the interaction between the cavity and mechanical modes can be very different, depending on the reflectivity of the mirror as well as its position relative to the wave nodes of the cavity mode. For a mirror (i.e., the membrane in the middle) with very low reflectivity, one can just consider a single cavity mode for the space between the HRM and the right cavity mirror, i.e., $j_{\text{max}}=1$ and $b=a$. In this case, if the middle mirror (assuming that its thickness is much smaller than the cavity field wavelength) is placed in the vicinity of a wave node (or anti-node), the optomechanical coupling between the cavity and mechanical modes is dominated by its \textit{quadratic} term, i.e., $H_{\text{om}}\propto b^{\dag}bx^{2}$ (similar for the interaction between $x$ and $a$), rather than its first-order term~\cite{Harris2008nature}. Otherwise, for a highly reflective mirror, mode $b$ serves as an independent cavity mode on the left side of the mirror, and the optomechanical couplings (between $x$ and both $a$ and $b$) typically have a linear dependence on the mechanical displacement.

\subsubsection{Non-Markovian reservoir regime}

Between the above two limits, the whole setup must show a continuous and smooth variation (rather than an abrupt transition) when changing the length $d$ of the loop~\cite{CWtransition}. For large enough (but not infinite) $d$ such that $\omega_{\text{FSR}}$ is much smaller than the coupling amplitude $\xi_{j}$, the cavity mode is coupled to \textit{a large number of} loop modes and $H_{\text{om}}$ can be neglected (this is justified also because the loop modes are not driven by pumping fields such that the corresponding optomechanical couplings cannot be enhanced effectively). In this case, the interaction part of the Hamiltonian becomes
\begin{equation}
H_{\text{int,non-Markov}}\simeq\sum_{j=1}^{j_{\text{max}}}\xi_{j}a^{\dag}\sin{(k_{j}d)}b_{j}+\text{H.c.}
\label{HSLnmk}
\end{equation}
However, due to the large $d$, the optomechanical system is equivalent to be coupled to a \textit{non-Markovian} reservior with \textit{time-delayed} coherent feedback~\cite{CWtransition}. For instance, for $d\sim 1\,\text{m}$ (i.e., $\omega_{\text{FSR}}\sim 10^{8}\,\text{Hz}$) and $\xi_{j}\sim 10^{10}\,\text{Hz}$ (i.e., $\kappa_{1}\sim 10^{12}\,\text{Hz}$), the cavity mode $a$ can interact with more than $10^{4}$ modes in the loop, but meanwhile the propagation time $\tau=2d/c$ is much larger than the lifetime $1/(2\kappa_{1}+\kappa_{2})$ of the cavity mode.

\subsection{Traveling-wave version}\label{sec_traveling}
In order to implement a single-sided coherent feedback loop with negligible time delay, one can consider a \textit{traveling-wave} version as shown in Fig.~\ref{Sloop}. This model is in fact equivalent to a side-coupled structure with two separate coupling points, which can be viewed as an optomechanical analogue of ``giant atoms''~\cite{fiveyear}. Since the traveling-wave field contains a dense continuum of modes by nature, one does not have to use a very long loop (now $d$ is the optical path between the two circulators) and thus the time delay can be negligible compared to the lifetime of the cavity. In this case, the interaction part of the Hamiltonian can be given by
\begin{equation}
H_{\text{int,travel}}\simeq\sum_{j=1}^{+\infty}\xi_{j}a^{\dag}\left(1+e^{ik_{j}d}\right)b_{j}+\text{H.c.},
\label{HSLnmk}
\end{equation}
where the function $[1+\text{exp}(ik_{j}d)]$ accounts for the ``two-time'' interaction between the system and the traveling-wave field. This corresponds to the input-output formalism presented in Sec.~\ref{secSga}, where the traveling field interacts with the system twice, while accumulating a phase difference.

The main difference between the direct-coupled and side-coupled structures is \textit{whether the feedback loop is part of the freely-propagating traveling-wave field}. This difference is very important since it determines whether the free spectral range of the loop modes is determined by the length (and thus the time delay) of the loop.\\

\section{Ground-state cooling without coupling between Fano and mechanical modes}\label{decouple}
\stepcounter{part}

\begin{figure}[h]
\centering
\includegraphics[width=8.5 cm]{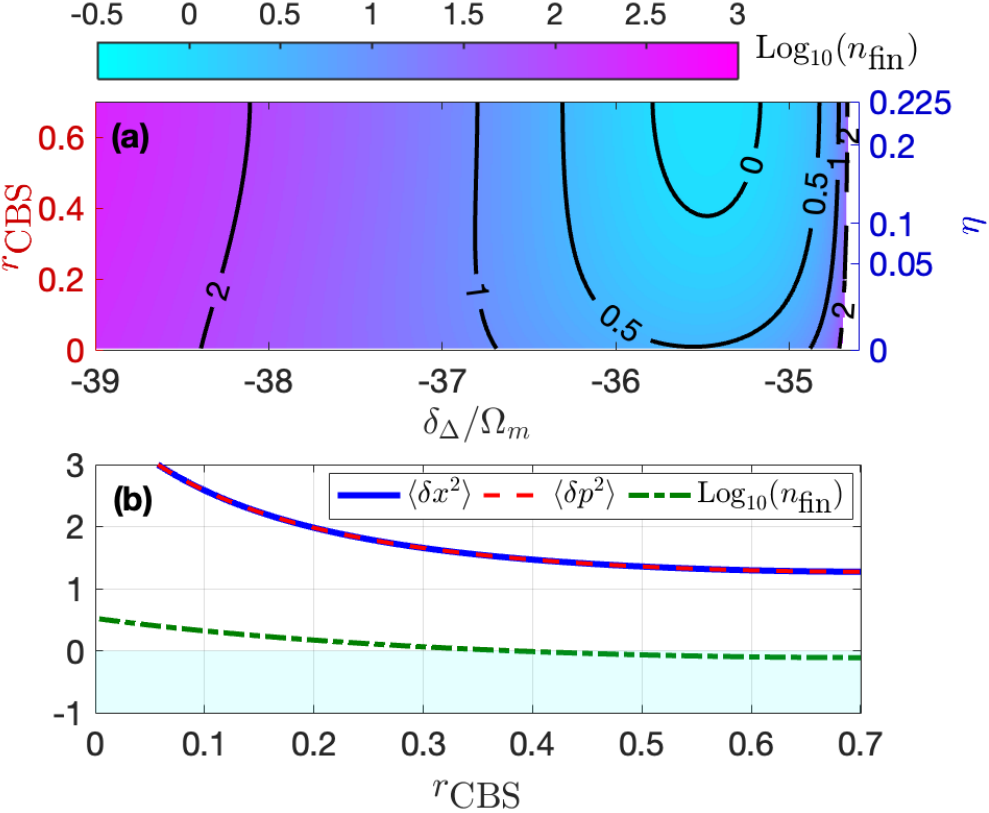}
\caption{(a) Final phonon number $n_\mathrm{fin}$ (on logarithmic scale) versus detuning $\delta_{\Delta}$ and reflection coefficient $r_\mathrm{CBS}$ for the setup with a single-sided coherent feedback but no Fano-mechanical coupling. (b) Final phonon number $n_\mathrm{fin}$ (on logarithmic scale) and mechanical variances ($\langle\delta x^{2}\rangle$ and $\langle\delta p^{2}\rangle$) versus $\delta_{\Delta}$ with $r_{\text{CBS}}=0.7$. Other parameters are identical to those in Fig.~\figpanel{SL1}{a} except for $\kappa_{1}/2\pi=30\,\text{THz}$, $g_{f,0}=0$, and $\varepsilon_{\text{p}}/2\pi=238.7\,\text{THz}$.}
\label{Nogf}
\end{figure}

As discussed in Sec.~\ref{secSgb}, the enhanced cooling effect, which is based on the combination of the Fano resonance and the coherent feedback, is not exclusive to the specific setup where the Fano and mechanical modes are coupled to each other via the deformation of the membrane. In fact, there are many different optomechanical setups where the Fano and mechanical modes do not directly interact with each other because, for example, they are supported by different objects~\cite{AE1,AE2,AE3,2cavity1,2cavity2,2cavity3,YCLiu2015pra} or the mechanical displacement only very weakly affects the properties of the Fano mode~\cite{GenesDoped2014}.

We provide in Fig.~\ref{Nogf} a proof-of-principle demonstration of ground-state cooling in this kind of setup. It shows that ground-state cooling is still allowed by resorting to the coherent feedback, with slightly modified parameters. The two mechanical variances $\langle\delta x^{2}\rangle$ and $\langle\delta p^{2}\rangle$ always show good agreement in the region of $n_{\text{fin}}<1$, ensuring the equipartition of energy. We thus conclude that ground-state cooling can also be achieved if there is no direct dispersive coupling between the Fano and the mechanical modes, but the mechanical mode is coupled to the cavity mode only.

\section{Comparison between Fano-mirror and coupled-cavity cooling scheme}\label{appCCS}
\stepcounter{part}

\begin{figure}[ptb]
\centering
\includegraphics[width=0.95\linewidth]{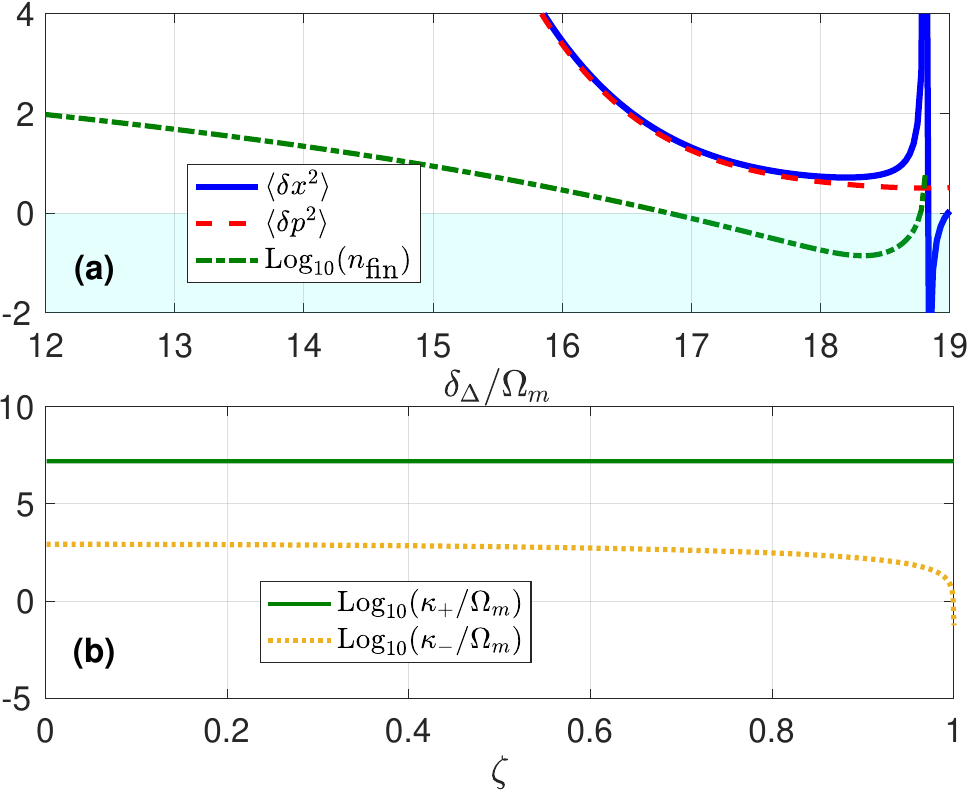}
\caption{(a) Final phonon number $n_{\text{fin}}$ (on logarithmic scale) and mechanical variances ($\langle\delta x^{2}\rangle$ and $\langle\delta p^{2}\rangle$) versus detuning $\delta_{\Delta}$ for our Fano-mirror setup, with parameters differing from those considered elsewhere and in the absence of the single-sided coherent feedback (i.e., $\eta=0$). (b) Effective linewidths $\kappa_{\pm}$ (on logarithmic scale) of the optical normal modes versus the dimensionless control parameter $\zeta$ for $\eta=0$. We take \blue{$\zeta=1$} in (a) and $\delta_{\Delta}=17.5$ in (b). The two panels share the same legend. Here we assume $\lambda/2\pi=0.9\,\text{GHz}$ and $\varepsilon_{\text{p}}/2\pi=0.8\,\text{THz}$. Other parameters are the same as those in Fig.~\figpanel{SL1}{a}.}\label{CoupledCavity}
\end{figure}

Fano resonances are one of the core ingredients in our cooling scheme. They have been extensively studied in a variety of optomechanical systems, including the coupled-cavity optomechanical setup where a sideband-unresolved optomechanical cavity is coupled to a high-quality bare cavity~\cite{YCLiu2015pra}.

Similar to the coupled-cavity cooling schemes, where ground-state cooling of the mechanical mode can be achieved by only resorting to the Fano resonance mechanism, in our Fano-mirror setup it is also possible to realize mechanical ground-state cooling without using the coherent feedback, but within a rather specific parametric regime, as discussed in Refs.~\cite{Juliette2021pra,Juliette2023arxiv}. We provide a specific demonstration in Fig.~\figpanel{CoupledCavity}{a} to support this conclusion.
While this strategy is theoretically viable, it has not yet been realized in experiments. In contrast, the single-sided feedback scheme proposed in this paper allows for robust ground-state cooling over a broad range of parameters, where it would otherwise be hindered by only exploiting the Fano resonance. For example, in the case of Fig.~\figpanel{CoupledCavity}{a}, ground-state cooling is realized with a stronger pumping field and a much weaker optical coherent coupling $\lambda$.

In this appendix, we would like to elaborate on the major differences between two important systems with Fano resonances, namely the coupled-cavity and the Fano-mirror optomechanical setups: (i) In the coupled-cavity setup, the Fano mode (i.e., the auxiliary cavity mode) is decoupled from the mechanical mode. (ii) In the coupled-cavity setup, the two optical modes are coupled solely through coherent interactions (without the optical dissipative coupling). Since we have studied in Appendix~\ref{decouple} the situation where the Fano mode does not interact with the mechanical mode, here instead we focus on the impact of the optical dissipative coupling on the Fano resonance (i.e., the optical normal modes). As will be shown below, in the presence of the optical dissipative coupling, the Fano resonance can facilitate ground-state cooling even if the linewidth of the Fano mode is much larger than the mechanical frequency.

In Fig.~\figpanel{CoupledCavity}{b}, we show the effective decay rates $\kappa_{\pm}$ of the two optical normal modes as a function of the dimensionless parameter $\zeta$, which controls the contribution of the optical dissipative coupling, defined by $G=\lambda-i\zeta\kappa_{1f}$. The optomechanical systems without such dissipative contributions, such as the coupled-cavity setup, can be captured by the case of $\zeta=0$. One can find that the optical dissipative coupling plays a crucial role in significantly reducing one of the effective decay rates (and thus making the system sideband-resolved as $\zeta$ approaches $1$). More specifically, the smaller decay rate can be reduced to well below the mechanical frequency $\Omega_{m}$ when $\zeta$ approaches $1$. This again explains why the single-sided coherent feedback, studied in the main text, plays a positive role when constructive interference occurs (e.g., $\phi=\pi$). 

\color{black}

\bibliography{OMfeedback}

\begin{thebibliography}{63}%
\makeatletter
\providecommand \@ifxundefined [1]{%
 \@ifx{#1\undefined}
}%
\providecommand \@ifnum [1]{%
 \ifnum #1\expandafter \@firstoftwo
 \else \expandafter \@secondoftwo
 \fi
}%
\providecommand \@ifx [1]{%
 \ifx #1\expandafter \@firstoftwo
 \else \expandafter \@secondoftwo
 \fi
}%
\providecommand \natexlab [1]{#1}%
\providecommand \enquote  [1]{``#1''}%
\providecommand \bibnamefont  [1]{#1}%
\providecommand \bibfnamefont [1]{#1}%
\providecommand \citenamefont [1]{#1}%
\providecommand \href@noop [0]{\@secondoftwo}%
\providecommand \href [0]{\begingroup \@sanitize@url \@href}%
\providecommand \@href[1]{\@@startlink{#1}\@@href}%
\providecommand \@@href[1]{\endgroup#1\@@endlink}%
\providecommand \@sanitize@url [0]{\catcode `\\12\catcode `\$12\catcode `\&12\catcode `\#12\catcode `\^12\catcode `\_12\catcode `\%12\relax}%
\providecommand \@@startlink[1]{}%
\providecommand \@@endlink[0]{}%
\providecommand \url  [0]{\begingroup\@sanitize@url \@url }%
\providecommand \@url [1]{\endgroup\@href {#1}{\urlprefix }}%
\providecommand \urlprefix  [0]{URL }%
\providecommand \Eprint [0]{\href }%
\providecommand \doibase [0]{https://doi.org/}%
\providecommand \selectlanguage [0]{\@gobble}%
\providecommand \bibinfo  [0]{\@secondoftwo}%
\providecommand \bibfield  [0]{\@secondoftwo}%
\providecommand \translation [1]{[#1]}%
\providecommand \BibitemOpen [0]{}%
\providecommand \bibitemStop [0]{}%
\providecommand \bibitemNoStop [0]{.\EOS\space}%
\providecommand \EOS [0]{\spacefactor3000\relax}%
\providecommand \BibitemShut  [1]{\csname bibitem#1\endcsname}%
\let\auto@bib@innerbib\@empty
\bibitem [{\citenamefont {Aspelmeyer}\ \emph {et~al.}(2014)\citenamefont {Aspelmeyer}, \citenamefont {Kippenberg},\ and\ \citenamefont {Marquardt}}]{OM2014RMP}%
  \BibitemOpen
  \bibfield  {author} {\bibinfo {author} {\bibfnamefont {M.}~\bibnamefont {Aspelmeyer}}, \bibinfo {author} {\bibfnamefont {T.~J.}\ \bibnamefont {Kippenberg}},\ and\ \bibinfo {author} {\bibfnamefont {F.}~\bibnamefont {Marquardt}},\ }\bibfield  {title} {\bibinfo {title} {Cavity optomechanics},\ }\href {https://doi.org/10.1103/RevModPhys.86.1391} {\bibfield  {journal} {\bibinfo  {journal} {Rev. Mod. Phys.}\ }\textbf {\bibinfo {volume} {86}},\ \bibinfo {pages} {1391} (\bibinfo {year} {2014})}\BibitemShut {NoStop}%
\bibitem [{\citenamefont {Regal}\ \emph {et~al.}(2008)\citenamefont {Regal}, \citenamefont {Teufel},\ and\ \citenamefont {Lehnert}}]{OMmeasure1}%
  \BibitemOpen
  \bibfield  {author} {\bibinfo {author} {\bibfnamefont {C.~A.}\ \bibnamefont {Regal}}, \bibinfo {author} {\bibfnamefont {J.~D.}\ \bibnamefont {Teufel}},\ and\ \bibinfo {author} {\bibfnamefont {K.~W.}\ \bibnamefont {Lehnert}},\ }\bibfield  {title} {\bibinfo {title} {Measuring nanomechanical motion with a microwave cavity interferometer},\ }\href {https://doi.org/10.1038/nphys974} {\bibfield  {journal} {\bibinfo  {journal} {Nat. Phys.}\ }\textbf {\bibinfo {volume} {4}},\ \bibinfo {pages} {555} (\bibinfo {year} {2008})}\BibitemShut {NoStop}%
\bibitem [{\citenamefont {Schliesser}\ \emph {et~al.}(2009)\citenamefont {Schliesser}, \citenamefont {Arcizet}, \citenamefont {Rivi{\`e}re}, \citenamefont {Anetsberger},\ and\ \citenamefont {Kippenberg}}]{OMmeasure2}%
  \BibitemOpen
  \bibfield  {author} {\bibinfo {author} {\bibfnamefont {A.}~\bibnamefont {Schliesser}}, \bibinfo {author} {\bibfnamefont {O.}~\bibnamefont {Arcizet}}, \bibinfo {author} {\bibfnamefont {R.}~\bibnamefont {Rivi{\`e}re}}, \bibinfo {author} {\bibfnamefont {G.}~\bibnamefont {Anetsberger}},\ and\ \bibinfo {author} {\bibfnamefont {T.~J.}\ \bibnamefont {Kippenberg}},\ }\bibfield  {title} {\bibinfo {title} {Resolved-sideband cooling and position measurement of a micromechanical oscillator close to the heisenberg uncertainty limit},\ }\href {https://doi.org/10.1038/nphys1304} {\bibfield  {journal} {\bibinfo  {journal} {Nat. Phys.}\ }\textbf {\bibinfo {volume} {5}},\ \bibinfo {pages} {509} (\bibinfo {year} {2009})}\BibitemShut {NoStop}%
\bibitem [{\citenamefont {Gavartin}\ \emph {et~al.}(2012)\citenamefont {Gavartin}, \citenamefont {Verlot},\ and\ \citenamefont {Kippenberg}}]{OMmeasure3}%
  \BibitemOpen
  \bibfield  {author} {\bibinfo {author} {\bibfnamefont {E.}~\bibnamefont {Gavartin}}, \bibinfo {author} {\bibfnamefont {P.}~\bibnamefont {Verlot}},\ and\ \bibinfo {author} {\bibfnamefont {T.~J.}\ \bibnamefont {Kippenberg}},\ }\bibfield  {title} {\bibinfo {title} {A hybrid on-chip optomechanical transducer for ultrasensitive force measurements},\ }\href {https://doi.org/10.1038/nnano.2012.97} {\bibfield  {journal} {\bibinfo  {journal} {Nat. Nanotechnol.}\ }\textbf {\bibinfo {volume} {7}},\ \bibinfo {pages} {509} (\bibinfo {year} {2012})}\BibitemShut {NoStop}%
\bibitem [{\citenamefont {Palomaki}\ \emph {et~al.}(2013)\citenamefont {Palomaki}, \citenamefont {Harlow}, \citenamefont {Teufel}, \citenamefont {Simmonds},\ and\ \citenamefont {Lehnert}}]{OMtransfer}%
  \BibitemOpen
  \bibfield  {author} {\bibinfo {author} {\bibfnamefont {T.}~\bibnamefont {Palomaki}}, \bibinfo {author} {\bibfnamefont {J.}~\bibnamefont {Harlow}}, \bibinfo {author} {\bibfnamefont {J.}~\bibnamefont {Teufel}}, \bibinfo {author} {\bibfnamefont {R.}~\bibnamefont {Simmonds}},\ and\ \bibinfo {author} {\bibfnamefont {K.~W.}\ \bibnamefont {Lehnert}},\ }\bibfield  {title} {\bibinfo {title} {Coherent state transfer between itinerant microwave fields and a mechanical oscillator},\ }\href {https://doi.org/10.1038/nature11915} {\bibfield  {journal} {\bibinfo  {journal} {Nature}\ }\textbf {\bibinfo {volume} {495}},\ \bibinfo {pages} {210} (\bibinfo {year} {2013})}\BibitemShut {NoStop}%
\bibitem [{\citenamefont {Dong}\ \emph {et~al.}(2012)\citenamefont {Dong}, \citenamefont {Fiore}, \citenamefont {Kuzyk},\ and\ \citenamefont {Wang}}]{Fconversion1}%
  \BibitemOpen
  \bibfield  {author} {\bibinfo {author} {\bibfnamefont {C.}~\bibnamefont {Dong}}, \bibinfo {author} {\bibfnamefont {V.}~\bibnamefont {Fiore}}, \bibinfo {author} {\bibfnamefont {M.~C.}\ \bibnamefont {Kuzyk}},\ and\ \bibinfo {author} {\bibfnamefont {H.}~\bibnamefont {Wang}},\ }\bibfield  {title} {\bibinfo {title} {Optomechanical dark mode},\ }\href {https://doi.org/10.1126/science.1228370} {\bibfield  {journal} {\bibinfo  {journal} {Science}\ }\textbf {\bibinfo {volume} {338}},\ \bibinfo {pages} {1609} (\bibinfo {year} {2012})}\BibitemShut {NoStop}%
\bibitem [{\citenamefont {Hill}\ \emph {et~al.}(2012)\citenamefont {Hill}, \citenamefont {Safavi-Naeini}, \citenamefont {Chan},\ and\ \citenamefont {Painter}}]{Fconversion2}%
  \BibitemOpen
  \bibfield  {author} {\bibinfo {author} {\bibfnamefont {J.~T.}\ \bibnamefont {Hill}}, \bibinfo {author} {\bibfnamefont {A.~H.}\ \bibnamefont {Safavi-Naeini}}, \bibinfo {author} {\bibfnamefont {J.}~\bibnamefont {Chan}},\ and\ \bibinfo {author} {\bibfnamefont {O.}~\bibnamefont {Painter}},\ }\bibfield  {title} {\bibinfo {title} {Coherent optical wavelength conversion via cavity optomechanics},\ }\href {https://doi.org/10.1038/ncomms2201} {\bibfield  {journal} {\bibinfo  {journal} {Nat. Commun.}\ }\textbf {\bibinfo {volume} {3}},\ \bibinfo {pages} {1196} (\bibinfo {year} {2012})}\BibitemShut {NoStop}%
\bibitem [{\citenamefont {Ockeloen-Korppi}\ \emph {et~al.}(2016)\citenamefont {Ockeloen-Korppi}, \citenamefont {Damsk\"agg}, \citenamefont {Pirkkalainen}, \citenamefont {Heikkil\"a}, \citenamefont {Massel},\ and\ \citenamefont {Sillanp\"a\"a}}]{Fconversion3}%
  \BibitemOpen
  \bibfield  {author} {\bibinfo {author} {\bibfnamefont {C.~F.}\ \bibnamefont {Ockeloen-Korppi}}, \bibinfo {author} {\bibfnamefont {E.}~\bibnamefont {Damsk\"agg}}, \bibinfo {author} {\bibfnamefont {J.-M.}\ \bibnamefont {Pirkkalainen}}, \bibinfo {author} {\bibfnamefont {T.~T.}\ \bibnamefont {Heikkil\"a}}, \bibinfo {author} {\bibfnamefont {F.}~\bibnamefont {Massel}},\ and\ \bibinfo {author} {\bibfnamefont {M.~A.}\ \bibnamefont {Sillanp\"a\"a}},\ }\bibfield  {title} {\bibinfo {title} {Low-noise amplification and frequency conversion with a multiport microwave optomechanical device},\ }\href {https://doi.org/10.1103/PhysRevX.6.041024} {\bibfield  {journal} {\bibinfo  {journal} {Phys. Rev. X}\ }\textbf {\bibinfo {volume} {6}},\ \bibinfo {pages} {041024} (\bibinfo {year} {2016})}\BibitemShut {NoStop}%
\bibitem [{\citenamefont {{\v{C}}ernot{\'\i}k}\ \emph {et~al.}(2018)\citenamefont {{\v{C}}ernot{\'\i}k}, \citenamefont {Mahmoodian},\ and\ \citenamefont {Hammerer}}]{Fconversion4}%
  \BibitemOpen
  \bibfield  {author} {\bibinfo {author} {\bibfnamefont {O.}~\bibnamefont {{\v{C}}ernot{\'\i}k}}, \bibinfo {author} {\bibfnamefont {S.}~\bibnamefont {Mahmoodian}},\ and\ \bibinfo {author} {\bibfnamefont {K.}~\bibnamefont {Hammerer}},\ }\bibfield  {title} {\bibinfo {title} {Spatially adiabatic frequency conversion in optoelectromechanical arrays},\ }\href {https://doi.org/10.1103/PhysRevLett.121.110506} {\bibfield  {journal} {\bibinfo  {journal} {Phys. Rev. Lett.}\ }\textbf {\bibinfo {volume} {121}},\ \bibinfo {pages} {110506} (\bibinfo {year} {2018})}\BibitemShut {NoStop}%
\bibitem [{\citenamefont {Kleckner}\ \emph {et~al.}(2008)\citenamefont {Kleckner}, \citenamefont {Pikovski}, \citenamefont {Jeffrey}, \citenamefont {Ament}, \citenamefont {Eliel}, \citenamefont {van~den Brink},\ and\ \citenamefont {Bouwmeester}}]{Ftest1}%
  \BibitemOpen
  \bibfield  {author} {\bibinfo {author} {\bibfnamefont {D.}~\bibnamefont {Kleckner}}, \bibinfo {author} {\bibfnamefont {I.}~\bibnamefont {Pikovski}}, \bibinfo {author} {\bibfnamefont {E.}~\bibnamefont {Jeffrey}}, \bibinfo {author} {\bibfnamefont {L.}~\bibnamefont {Ament}}, \bibinfo {author} {\bibfnamefont {E.}~\bibnamefont {Eliel}}, \bibinfo {author} {\bibfnamefont {J.}~\bibnamefont {van~den Brink}},\ and\ \bibinfo {author} {\bibfnamefont {D.}~\bibnamefont {Bouwmeester}},\ }\bibfield  {title} {\bibinfo {title} {Creating and verifying a quantum superposition in a micro-optomechanical system},\ }\href {https://doi.org/10.1088/1367-2630/10/9/095020} {\bibfield  {journal} {\bibinfo  {journal} {New J. Phys.}\ }\textbf {\bibinfo {volume} {10}},\ \bibinfo {pages} {095020} (\bibinfo {year} {2008})}\BibitemShut {NoStop}%
\bibitem [{\citenamefont {Romero-Isart}\ \emph {et~al.}(2011)\citenamefont {Romero-Isart}, \citenamefont {Pflanzer}, \citenamefont {Blaser}, \citenamefont {Kaltenbaek}, \citenamefont {Kiesel}, \citenamefont {Aspelmeyer},\ and\ \citenamefont {Cirac}}]{Ftest2}%
  \BibitemOpen
  \bibfield  {author} {\bibinfo {author} {\bibfnamefont {O.}~\bibnamefont {Romero-Isart}}, \bibinfo {author} {\bibfnamefont {A.~C.}\ \bibnamefont {Pflanzer}}, \bibinfo {author} {\bibfnamefont {F.}~\bibnamefont {Blaser}}, \bibinfo {author} {\bibfnamefont {R.}~\bibnamefont {Kaltenbaek}}, \bibinfo {author} {\bibfnamefont {N.}~\bibnamefont {Kiesel}}, \bibinfo {author} {\bibfnamefont {M.}~\bibnamefont {Aspelmeyer}},\ and\ \bibinfo {author} {\bibfnamefont {J.~I.}\ \bibnamefont {Cirac}},\ }\bibfield  {title} {\bibinfo {title} {Large quantum superpositions and interference of massive nanometer-sized objects},\ }\href {https://doi.org/10.1103/PhysRevLett.107.020405} {\bibfield  {journal} {\bibinfo  {journal} {Phys. Rev. Lett.}\ }\textbf {\bibinfo {volume} {107}},\ \bibinfo {pages} {020405} (\bibinfo {year} {2011})}\BibitemShut {NoStop}%
\bibitem [{\citenamefont {Pikovski}\ \emph {et~al.}(2012)\citenamefont {Pikovski}, \citenamefont {Vanner}, \citenamefont {Aspelmeyer}, \citenamefont {Kim},\ and\ \citenamefont {Brukner}}]{Ftest3}%
  \BibitemOpen
  \bibfield  {author} {\bibinfo {author} {\bibfnamefont {I.}~\bibnamefont {Pikovski}}, \bibinfo {author} {\bibfnamefont {M.~R.}\ \bibnamefont {Vanner}}, \bibinfo {author} {\bibfnamefont {M.}~\bibnamefont {Aspelmeyer}}, \bibinfo {author} {\bibfnamefont {M.}~\bibnamefont {Kim}},\ and\ \bibinfo {author} {\bibfnamefont {{\v{C}}.}~\bibnamefont {Brukner}},\ }\bibfield  {title} {\bibinfo {title} {Probing planck-scale physics with quantum optics},\ }\href {https://doi.org/10.1038/nphys2262} {\bibfield  {journal} {\bibinfo  {journal} {Nat. Phys.}\ }\textbf {\bibinfo {volume} {8}},\ \bibinfo {pages} {393} (\bibinfo {year} {2012})}\BibitemShut {NoStop}%
\bibitem [{\citenamefont {Barzanjeh}\ \emph {et~al.}(2022)\citenamefont {Barzanjeh}, \citenamefont {Xuereb}, \citenamefont {Gr{\"o}blacher}, \citenamefont {Paternostro}, \citenamefont {Regal},\ and\ \citenamefont {Weig}}]{OMQT}%
  \BibitemOpen
  \bibfield  {author} {\bibinfo {author} {\bibfnamefont {S.}~\bibnamefont {Barzanjeh}}, \bibinfo {author} {\bibfnamefont {A.}~\bibnamefont {Xuereb}}, \bibinfo {author} {\bibfnamefont {S.}~\bibnamefont {Gr{\"o}blacher}}, \bibinfo {author} {\bibfnamefont {M.}~\bibnamefont {Paternostro}}, \bibinfo {author} {\bibfnamefont {C.~A.}\ \bibnamefont {Regal}},\ and\ \bibinfo {author} {\bibfnamefont {E.~M.}\ \bibnamefont {Weig}},\ }\bibfield  {title} {\bibinfo {title} {Optomechanics for quantum technologies},\ }\href {https://doi.org/10.1038/s41567-021-01402-0} {\bibfield  {journal} {\bibinfo  {journal} {Nat. Phys.}\ }\textbf {\bibinfo {volume} {18}},\ \bibinfo {pages} {15} (\bibinfo {year} {2022})}\BibitemShut {NoStop}%
\bibitem [{\citenamefont {Chan}\ \emph {et~al.}(2011)\citenamefont {Chan}, \citenamefont {Alegre}, \citenamefont {Safavi-Naeini}, \citenamefont {Hill}, \citenamefont {Krause}, \citenamefont {Gr{\"o}blacher}, \citenamefont {Aspelmeyer},\ and\ \citenamefont {Painter}}]{MBF4}%
  \BibitemOpen
  \bibfield  {author} {\bibinfo {author} {\bibfnamefont {J.}~\bibnamefont {Chan}}, \bibinfo {author} {\bibfnamefont {T.~M.}\ \bibnamefont {Alegre}}, \bibinfo {author} {\bibfnamefont {A.~H.}\ \bibnamefont {Safavi-Naeini}}, \bibinfo {author} {\bibfnamefont {J.~T.}\ \bibnamefont {Hill}}, \bibinfo {author} {\bibfnamefont {A.}~\bibnamefont {Krause}}, \bibinfo {author} {\bibfnamefont {S.}~\bibnamefont {Gr{\"o}blacher}}, \bibinfo {author} {\bibfnamefont {M.}~\bibnamefont {Aspelmeyer}},\ and\ \bibinfo {author} {\bibfnamefont {O.}~\bibnamefont {Painter}},\ }\bibfield  {title} {\bibinfo {title} {Laser cooling of a nanomechanical oscillator into its quantum ground state},\ }\href {https://doi.org/10.1038/nature10461} {\bibfield  {journal} {\bibinfo  {journal} {Nature}\ }\textbf {\bibinfo {volume} {478}},\ \bibinfo {pages} {89} (\bibinfo {year} {2011})}\BibitemShut {NoStop}%
\bibitem [{\citenamefont {Teufel}\ \emph {et~al.}(2011)\citenamefont {Teufel}, \citenamefont {Donner}, \citenamefont {Li}, \citenamefont {Harlow}, \citenamefont {Allman}, \citenamefont {Cicak}, \citenamefont {Sirois}, \citenamefont {Whittaker}, \citenamefont {Lehnert},\ and\ \citenamefont {Simmonds}}]{Teufel2011Jul}%
  \BibitemOpen
  \bibfield  {author} {\bibinfo {author} {\bibfnamefont {J.~D.}\ \bibnamefont {Teufel}}, \bibinfo {author} {\bibfnamefont {T.}~\bibnamefont {Donner}}, \bibinfo {author} {\bibfnamefont {D.}~\bibnamefont {Li}}, \bibinfo {author} {\bibfnamefont {J.~W.}\ \bibnamefont {Harlow}}, \bibinfo {author} {\bibfnamefont {M.~S.}\ \bibnamefont {Allman}}, \bibinfo {author} {\bibfnamefont {K.}~\bibnamefont {Cicak}}, \bibinfo {author} {\bibfnamefont {A.~J.}\ \bibnamefont {Sirois}}, \bibinfo {author} {\bibfnamefont {J.~D.}\ \bibnamefont {Whittaker}}, \bibinfo {author} {\bibfnamefont {K.~W.}\ \bibnamefont {Lehnert}},\ and\ \bibinfo {author} {\bibfnamefont {R.~W.}\ \bibnamefont {Simmonds}},\ }\bibfield  {title} {\bibinfo {title} {Sideband cooling of micromechanical motion to the quantum ground state},\ }\href {https://doi.org/10.1038/nature10261} {\bibfield  {journal} {\bibinfo  {journal} {Nature}\ }\textbf {\bibinfo {volume} {475}},\ \bibinfo {pages} {359} (\bibinfo {year} {2011})}\BibitemShut {NoStop}%
\bibitem [{\citenamefont {Deli{\'c}}\ \emph {et~al.}(2020)\citenamefont {Deli{\'c}}, \citenamefont {Reisenbauer}, \citenamefont {Dare}, \citenamefont {Grass}, \citenamefont {Vuleti{\'c}}, \citenamefont {Kiesel},\ and\ \citenamefont {Aspelmeyer}}]{Delic2020Feb}%
  \BibitemOpen
  \bibfield  {author} {\bibinfo {author} {\bibfnamefont {U.}~\bibnamefont {Deli{\'c}}}, \bibinfo {author} {\bibfnamefont {M.}~\bibnamefont {Reisenbauer}}, \bibinfo {author} {\bibfnamefont {K.}~\bibnamefont {Dare}}, \bibinfo {author} {\bibfnamefont {D.}~\bibnamefont {Grass}}, \bibinfo {author} {\bibfnamefont {V.}~\bibnamefont {Vuleti{\'c}}}, \bibinfo {author} {\bibfnamefont {N.}~\bibnamefont {Kiesel}},\ and\ \bibinfo {author} {\bibfnamefont {M.}~\bibnamefont {Aspelmeyer}},\ }\bibfield  {title} {\bibinfo {title} {Cooling of a levitated nanoparticle to the motional quantum ground state},\ }\href {https://doi.org/10.1126/science.aba3993} {\bibfield  {journal} {\bibinfo  {journal} {Science}\ }\textbf {\bibinfo {volume} {367}},\ \bibinfo {pages} {892} (\bibinfo {year} {2020})}\BibitemShut {NoStop}%
\bibitem [{\citenamefont {Zhang}\ \emph {et~al.}(2017)\citenamefont {Zhang}, \citenamefont {Liu}, \citenamefont {Wu}, \citenamefont {Jacobs},\ and\ \citenamefont {Nori}}]{ZhangJreview}%
  \BibitemOpen
  \bibfield  {author} {\bibinfo {author} {\bibfnamefont {J.}~\bibnamefont {Zhang}}, \bibinfo {author} {\bibfnamefont {Y.-X.}\ \bibnamefont {Liu}}, \bibinfo {author} {\bibfnamefont {R.-B.}\ \bibnamefont {Wu}}, \bibinfo {author} {\bibfnamefont {K.}~\bibnamefont {Jacobs}},\ and\ \bibinfo {author} {\bibfnamefont {F.}~\bibnamefont {Nori}},\ }\bibfield  {title} {\bibinfo {title} {Quantum feedback: Theory, experiments, and applications},\ }\href {https://doi.org/https://doi.org/10.1016/j.physrep.2017.02.003} {\bibfield  {journal} {\bibinfo  {journal} {Phys. Rep.}\ }\textbf {\bibinfo {volume} {679}},\ \bibinfo {pages} {1} (\bibinfo {year} {2017})}\BibitemShut {NoStop}%
\bibitem [{\citenamefont {Sommer}\ and\ \citenamefont {Genes}(2019)}]{MBF1}%
  \BibitemOpen
  \bibfield  {author} {\bibinfo {author} {\bibfnamefont {C.}~\bibnamefont {Sommer}}\ and\ \bibinfo {author} {\bibfnamefont {C.}~\bibnamefont {Genes}},\ }\bibfield  {title} {\bibinfo {title} {Partial optomechanical refrigeration via multimode cold-damping feedback},\ }\href {https://doi.org/10.1103/PhysRevLett.123.203605} {\bibfield  {journal} {\bibinfo  {journal} {Phys. Rev. Lett.}\ }\textbf {\bibinfo {volume} {123}},\ \bibinfo {pages} {203605} (\bibinfo {year} {2019})}\BibitemShut {NoStop}%
\bibitem [{\citenamefont {Sommer}\ \emph {et~al.}(2020)\citenamefont {Sommer}, \citenamefont {Ghosh},\ and\ \citenamefont {Genes}}]{MBF2}%
  \BibitemOpen
  \bibfield  {author} {\bibinfo {author} {\bibfnamefont {C.}~\bibnamefont {Sommer}}, \bibinfo {author} {\bibfnamefont {A.}~\bibnamefont {Ghosh}},\ and\ \bibinfo {author} {\bibfnamefont {C.}~\bibnamefont {Genes}},\ }\bibfield  {title} {\bibinfo {title} {Multimode cold-damping optomechanics with delayed feedback},\ }\href {https://doi.org/10.1103/PhysRevResearch.2.033299} {\bibfield  {journal} {\bibinfo  {journal} {Phys. Rev. Res.}\ }\textbf {\bibinfo {volume} {2}},\ \bibinfo {pages} {033299} (\bibinfo {year} {2020})}\BibitemShut {NoStop}%
\bibitem [{\citenamefont {Lai}\ \emph {et~al.}(2021)\citenamefont {Lai}, \citenamefont {Huang}, \citenamefont {Hou}, \citenamefont {Nori},\ and\ \citenamefont {Liao}}]{MBF3}%
  \BibitemOpen
  \bibfield  {author} {\bibinfo {author} {\bibfnamefont {D.-G.}\ \bibnamefont {Lai}}, \bibinfo {author} {\bibfnamefont {J.}~\bibnamefont {Huang}}, \bibinfo {author} {\bibfnamefont {B.-P.}\ \bibnamefont {Hou}}, \bibinfo {author} {\bibfnamefont {F.}~\bibnamefont {Nori}},\ and\ \bibinfo {author} {\bibfnamefont {J.-Q.}\ \bibnamefont {Liao}},\ }\bibfield  {title} {\bibinfo {title} {Domino cooling of a coupled mechanical-resonator chain via cold-damping feedback},\ }\href {https://doi.org/10.1103/PhysRevA.103.063509} {\bibfield  {journal} {\bibinfo  {journal} {Phys. Rev. A}\ }\textbf {\bibinfo {volume} {103}},\ \bibinfo {pages} {063509} (\bibinfo {year} {2021})}\BibitemShut {NoStop}%
\bibitem [{\citenamefont {Rossi}\ \emph {et~al.}(2018)\citenamefont {Rossi}, \citenamefont {Mason}, \citenamefont {Chen}, \citenamefont {Tsaturyan},\ and\ \citenamefont {Schliesser}}]{MBF5}%
  \BibitemOpen
  \bibfield  {author} {\bibinfo {author} {\bibfnamefont {M.}~\bibnamefont {Rossi}}, \bibinfo {author} {\bibfnamefont {D.}~\bibnamefont {Mason}}, \bibinfo {author} {\bibfnamefont {J.}~\bibnamefont {Chen}}, \bibinfo {author} {\bibfnamefont {Y.}~\bibnamefont {Tsaturyan}},\ and\ \bibinfo {author} {\bibfnamefont {A.}~\bibnamefont {Schliesser}},\ }\bibfield  {title} {\bibinfo {title} {Measurement-based quantum control of mechanical motion},\ }\href {https://doi.org/10.1038/s41586-018-0643-8} {\bibfield  {journal} {\bibinfo  {journal} {Nature}\ }\textbf {\bibinfo {volume} {563}},\ \bibinfo {pages} {53} (\bibinfo {year} {2018})}\BibitemShut {NoStop}%
\bibitem [{\citenamefont {Qiu}\ \emph {et~al.}(2022)\citenamefont {Qiu}, \citenamefont {Huang}, \citenamefont {Shomroni}, \citenamefont {Pan}, \citenamefont {Seidler},\ and\ \citenamefont {Kippenberg}}]{MBF6}%
  \BibitemOpen
  \bibfield  {author} {\bibinfo {author} {\bibfnamefont {L.}~\bibnamefont {Qiu}}, \bibinfo {author} {\bibfnamefont {G.}~\bibnamefont {Huang}}, \bibinfo {author} {\bibfnamefont {I.}~\bibnamefont {Shomroni}}, \bibinfo {author} {\bibfnamefont {J.}~\bibnamefont {Pan}}, \bibinfo {author} {\bibfnamefont {P.}~\bibnamefont {Seidler}},\ and\ \bibinfo {author} {\bibfnamefont {T.~J.}\ \bibnamefont {Kippenberg}},\ }\bibfield  {title} {\bibinfo {title} {Dissipative quantum feedback in measurements using a parametrically coupled microcavity},\ }\href {https://doi.org/10.1103/PRXQuantum.3.020309} {\bibfield  {journal} {\bibinfo  {journal} {PRX Quantum}\ }\textbf {\bibinfo {volume} {3}},\ \bibinfo {pages} {020309} (\bibinfo {year} {2022})}\BibitemShut {NoStop}%
\bibitem [{\citenamefont {Magrini}\ \emph {et~al.}(2021)\citenamefont {Magrini}, \citenamefont {Rosenzweig}, \citenamefont {Bach}, \citenamefont {Deutschmann-Olek}, \citenamefont {Hofer}, \citenamefont {Hong}, \citenamefont {Kiesel}, \citenamefont {Kugi},\ and\ \citenamefont {Aspelmeyer}}]{Magrini2021Jul}%
  \BibitemOpen
  \bibfield  {author} {\bibinfo {author} {\bibfnamefont {L.}~\bibnamefont {Magrini}}, \bibinfo {author} {\bibfnamefont {P.}~\bibnamefont {Rosenzweig}}, \bibinfo {author} {\bibfnamefont {C.}~\bibnamefont {Bach}}, \bibinfo {author} {\bibfnamefont {A.}~\bibnamefont {Deutschmann-Olek}}, \bibinfo {author} {\bibfnamefont {S.~G.}\ \bibnamefont {Hofer}}, \bibinfo {author} {\bibfnamefont {S.}~\bibnamefont {Hong}}, \bibinfo {author} {\bibfnamefont {N.}~\bibnamefont {Kiesel}}, \bibinfo {author} {\bibfnamefont {A.}~\bibnamefont {Kugi}},\ and\ \bibinfo {author} {\bibfnamefont {M.}~\bibnamefont {Aspelmeyer}},\ }\bibfield  {title} {\bibinfo {title} {{Real-time optimal quantum control of mechanical motion at room temperature}},\ }\href {https://doi.org/10.1038/s41586-021-03602-3} {\bibfield  {journal} {\bibinfo  {journal} {Nature}\ }\textbf {\bibinfo {volume} {595}},\ \bibinfo {pages} {373} (\bibinfo {year} {2021})}\BibitemShut {NoStop}%
\bibitem [{\citenamefont {Tebbenjohanns}\ \emph {et~al.}(2021)\citenamefont {Tebbenjohanns}, \citenamefont {Mattana}, \citenamefont {Rossi}, \citenamefont {Frimmer},\ and\ \citenamefont {Novotny}}]{Tebbenjohanns2021Jul}%
  \BibitemOpen
  \bibfield  {author} {\bibinfo {author} {\bibfnamefont {F.}~\bibnamefont {Tebbenjohanns}}, \bibinfo {author} {\bibfnamefont {M.~L.}\ \bibnamefont {Mattana}}, \bibinfo {author} {\bibfnamefont {M.}~\bibnamefont {Rossi}}, \bibinfo {author} {\bibfnamefont {M.}~\bibnamefont {Frimmer}},\ and\ \bibinfo {author} {\bibfnamefont {L.}~\bibnamefont {Novotny}},\ }\bibfield  {title} {\bibinfo {title} {{Quantum control of a nanoparticle optically levitated in cryogenic free space}},\ }\href {https://doi.org/10.1038/s41586-021-03617-w} {\bibfield  {journal} {\bibinfo  {journal} {Nature}\ }\textbf {\bibinfo {volume} {595}},\ \bibinfo {pages} {378} (\bibinfo {year} {2021})}\BibitemShut {NoStop}%
\bibitem [{\citenamefont {Manikandan}\ and\ \citenamefont {Qvarfort}(2023)}]{Manikandan2023Feb}%
  \BibitemOpen
  \bibfield  {author} {\bibinfo {author} {\bibfnamefont {S.~K.}\ \bibnamefont {Manikandan}}\ and\ \bibinfo {author} {\bibfnamefont {S.}~\bibnamefont {Qvarfort}},\ }\bibfield  {title} {\bibinfo {title} {Optimal quantum parametric feedback cooling},\ }\href {https://doi.org/10.1103/PhysRevA.107.023516} {\bibfield  {journal} {\bibinfo  {journal} {Physical Review A}\ }\textbf {\bibinfo {volume} {107}},\ \bibinfo {pages} {023516} (\bibinfo {year} {2023})}\BibitemShut {NoStop}%
\bibitem [{\citenamefont {Lloyd}(2000)}]{Lloyd2000}%
  \BibitemOpen
  \bibfield  {author} {\bibinfo {author} {\bibfnamefont {S.}~\bibnamefont {Lloyd}},\ }\bibfield  {title} {\bibinfo {title} {Coherent quantum feedback},\ }\href {https://doi.org/10.1103/PhysRevA.62.022108} {\bibfield  {journal} {\bibinfo  {journal} {Phys. Rev. A}\ }\textbf {\bibinfo {volume} {62}},\ \bibinfo {pages} {022108} (\bibinfo {year} {2000})}\BibitemShut {NoStop}%
\bibitem [{\citenamefont {Li}\ \emph {et~al.}(2017)\citenamefont {Li}, \citenamefont {Li}, \citenamefont {Zippilli}, \citenamefont {Vitali},\ and\ \citenamefont {Zhang}}]{LiJ2017}%
  \BibitemOpen
  \bibfield  {author} {\bibinfo {author} {\bibfnamefont {J.}~\bibnamefont {Li}}, \bibinfo {author} {\bibfnamefont {G.}~\bibnamefont {Li}}, \bibinfo {author} {\bibfnamefont {S.}~\bibnamefont {Zippilli}}, \bibinfo {author} {\bibfnamefont {D.}~\bibnamefont {Vitali}},\ and\ \bibinfo {author} {\bibfnamefont {T.}~\bibnamefont {Zhang}},\ }\bibfield  {title} {\bibinfo {title} {Enhanced entanglement of two different mechanical resonators via coherent feedback},\ }\href {https://doi.org/10.1103/PhysRevA.95.043819} {\bibfield  {journal} {\bibinfo  {journal} {Phys. Rev. A}\ }\textbf {\bibinfo {volume} {95}},\ \bibinfo {pages} {043819} (\bibinfo {year} {2017})}\BibitemShut {NoStop}%
\bibitem [{\citenamefont {Harwood}\ \emph {et~al.}(2021)\citenamefont {Harwood}, \citenamefont {Brunelli},\ and\ \citenamefont {Serafini}}]{Harwood2021}%
  \BibitemOpen
  \bibfield  {author} {\bibinfo {author} {\bibfnamefont {A.}~\bibnamefont {Harwood}}, \bibinfo {author} {\bibfnamefont {M.}~\bibnamefont {Brunelli}},\ and\ \bibinfo {author} {\bibfnamefont {A.}~\bibnamefont {Serafini}},\ }\bibfield  {title} {\bibinfo {title} {Cavity optomechanics assisted by optical coherent feedback},\ }\href {https://doi.org/10.1103/PhysRevA.103.023509} {\bibfield  {journal} {\bibinfo  {journal} {Phys. Rev. A}\ }\textbf {\bibinfo {volume} {103}},\ \bibinfo {pages} {023509} (\bibinfo {year} {2021})}\BibitemShut {NoStop}%
\bibitem [{\citenamefont {Huang}\ and\ \citenamefont {Chen}(2019)}]{HuangMPDI}%
  \BibitemOpen
  \bibfield  {author} {\bibinfo {author} {\bibfnamefont {S.}~\bibnamefont {Huang}}\ and\ \bibinfo {author} {\bibfnamefont {A.}~\bibnamefont {Chen}},\ }\bibfield  {title} {\bibinfo {title} {{Cooling of a Mechanical Oscillator and Normal Mode Splitting in Optomechanical Systems with Coherent Feedback}},\ }\href {https://doi.org/10.3390/app9163402} {\bibfield  {journal} {\bibinfo  {journal} {Appl. Sci.}\ }\textbf {\bibinfo {volume} {9}},\ \bibinfo {pages} {3402} (\bibinfo {year} {2019})}\BibitemShut {NoStop}%
\bibitem [{\citenamefont {Peng}\ \emph {et~al.}(2023)\citenamefont {Peng}, \citenamefont {Zhao}, \citenamefont {Yang}, \citenamefont {Yang},\ and\ \citenamefont {Zhou}}]{FaradayR}%
  \BibitemOpen
  \bibfield  {author} {\bibinfo {author} {\bibfnamefont {R.}~\bibnamefont {Peng}}, \bibinfo {author} {\bibfnamefont {C.}~\bibnamefont {Zhao}}, \bibinfo {author} {\bibfnamefont {Z.}~\bibnamefont {Yang}}, \bibinfo {author} {\bibfnamefont {J.}~\bibnamefont {Yang}},\ and\ \bibinfo {author} {\bibfnamefont {L.}~\bibnamefont {Zhou}},\ }\bibfield  {title} {\bibinfo {title} {Enhancement of mechanical entanglement and asymmetric steering with coherent feedback},\ }\href {https://doi.org/10.1103/PhysRevA.107.013507} {\bibfield  {journal} {\bibinfo  {journal} {Phys. Rev. A}\ }\textbf {\bibinfo {volume} {107}},\ \bibinfo {pages} {013507} (\bibinfo {year} {2023})}\BibitemShut {NoStop}%
\bibitem [{\citenamefont {Guo}\ and\ \citenamefont {Gr{\"{o}}blacher}(2022)}]{JKGuo}%
  \BibitemOpen
  \bibfield  {author} {\bibinfo {author} {\bibfnamefont {J.}~\bibnamefont {Guo}}\ and\ \bibinfo {author} {\bibfnamefont {S.}~\bibnamefont {Gr{\"{o}}blacher}},\ }\bibfield  {title} {\bibinfo {title} {Coherent feedback in optomechanical systems in the sideband-unresolved regime},\ }\href {https://doi.org/10.22331/q-2022-11-03-848} {\bibfield  {journal} {\bibinfo  {journal} {{Quantum}}\ }\textbf {\bibinfo {volume} {6}},\ \bibinfo {pages} {848} (\bibinfo {year} {2022})}\BibitemShut {NoStop}%
\bibitem [{\citenamefont {Ernzer}\ \emph {et~al.}(2023)\citenamefont {Ernzer}, \citenamefont {Bosch~Aguilera}, \citenamefont {Brunelli}, \citenamefont {Schmid}, \citenamefont {Karg}, \citenamefont {Bruder}, \citenamefont {Potts},\ and\ \citenamefont {Treutlein}}]{BaselPRX2023}%
  \BibitemOpen
  \bibfield  {author} {\bibinfo {author} {\bibfnamefont {M.}~\bibnamefont {Ernzer}}, \bibinfo {author} {\bibfnamefont {M.}~\bibnamefont {Bosch~Aguilera}}, \bibinfo {author} {\bibfnamefont {M.}~\bibnamefont {Brunelli}}, \bibinfo {author} {\bibfnamefont {G.-L.}\ \bibnamefont {Schmid}}, \bibinfo {author} {\bibfnamefont {T.~M.}\ \bibnamefont {Karg}}, \bibinfo {author} {\bibfnamefont {C.}~\bibnamefont {Bruder}}, \bibinfo {author} {\bibfnamefont {P.~P.}\ \bibnamefont {Potts}},\ and\ \bibinfo {author} {\bibfnamefont {P.}~\bibnamefont {Treutlein}},\ }\bibfield  {title} {\bibinfo {title} {Optical coherent feedback control of a mechanical oscillator},\ }\href {https://doi.org/10.1103/PhysRevX.13.021023} {\bibfield  {journal} {\bibinfo  {journal} {Phys. Rev. X}\ }\textbf {\bibinfo {volume} {13}},\ \bibinfo {pages} {021023} (\bibinfo {year} {2023})}\BibitemShut {NoStop}%
\bibitem [{\citenamefont {Limonov}\ \emph {et~al.}(2017)\citenamefont {Limonov}, \citenamefont {Rybin}, \citenamefont {Poddubny},\ and\ \citenamefont {Kivshar}}]{Limonov2017Sep}%
  \BibitemOpen
  \bibfield  {author} {\bibinfo {author} {\bibfnamefont {M.~F.}\ \bibnamefont {Limonov}}, \bibinfo {author} {\bibfnamefont {M.~V.}\ \bibnamefont {Rybin}}, \bibinfo {author} {\bibfnamefont {A.~N.}\ \bibnamefont {Poddubny}},\ and\ \bibinfo {author} {\bibfnamefont {Y.~S.}\ \bibnamefont {Kivshar}},\ }\bibfield  {title} {\bibinfo {title} {Fano resonances in photonics},\ }\href {https://doi.org/10.1038/nphoton.2017.142} {\bibfield  {journal} {\bibinfo  {journal} {Nature Photonics}\ }\textbf {\bibinfo {volume} {11}},\ \bibinfo {pages} {543} (\bibinfo {year} {2017})}\BibitemShut {NoStop}%
\bibitem [{\citenamefont {Genes}\ \emph {et~al.}(2011)\citenamefont {Genes}, \citenamefont {Ritsch}, \citenamefont {Drewsen},\ and\ \citenamefont {Dantan}}]{AE1}%
  \BibitemOpen
  \bibfield  {author} {\bibinfo {author} {\bibfnamefont {C.}~\bibnamefont {Genes}}, \bibinfo {author} {\bibfnamefont {H.}~\bibnamefont {Ritsch}}, \bibinfo {author} {\bibfnamefont {M.}~\bibnamefont {Drewsen}},\ and\ \bibinfo {author} {\bibfnamefont {A.}~\bibnamefont {Dantan}},\ }\bibfield  {title} {\bibinfo {title} {Atom-membrane cooling and entanglement using cavity electromagnetically induced transparency},\ }\href {https://doi.org/10.1103/PhysRevA.84.051801} {\bibfield  {journal} {\bibinfo  {journal} {Phys. Rev. A}\ }\textbf {\bibinfo {volume} {84}},\ \bibinfo {pages} {051801} (\bibinfo {year} {2011})}\BibitemShut {NoStop}%
\bibitem [{\citenamefont {Bariani}\ \emph {et~al.}(2014)\citenamefont {Bariani}, \citenamefont {Singh}, \citenamefont {Buchmann}, \citenamefont {Vengalattore},\ and\ \citenamefont {Meystre}}]{AE2}%
  \BibitemOpen
  \bibfield  {author} {\bibinfo {author} {\bibfnamefont {F.}~\bibnamefont {Bariani}}, \bibinfo {author} {\bibfnamefont {S.}~\bibnamefont {Singh}}, \bibinfo {author} {\bibfnamefont {L.~F.}\ \bibnamefont {Buchmann}}, \bibinfo {author} {\bibfnamefont {M.}~\bibnamefont {Vengalattore}},\ and\ \bibinfo {author} {\bibfnamefont {P.}~\bibnamefont {Meystre}},\ }\bibfield  {title} {\bibinfo {title} {Hybrid optomechanical cooling by atomic $\ensuremath{\Lambda}$ systems},\ }\href {https://doi.org/10.1103/PhysRevA.90.033838} {\bibfield  {journal} {\bibinfo  {journal} {Phys. Rev. A}\ }\textbf {\bibinfo {volume} {90}},\ \bibinfo {pages} {033838} (\bibinfo {year} {2014})}\BibitemShut {NoStop}%
\bibitem [{\citenamefont {J{\"o}ckel}\ \emph {et~al.}(2015)\citenamefont {J{\"o}ckel}, \citenamefont {Faber}, \citenamefont {Kampschulte}, \citenamefont {Korppi}, \citenamefont {Rakher},\ and\ \citenamefont {Treutlein}}]{AE3}%
  \BibitemOpen
  \bibfield  {author} {\bibinfo {author} {\bibfnamefont {A.}~\bibnamefont {J{\"o}ckel}}, \bibinfo {author} {\bibfnamefont {A.}~\bibnamefont {Faber}}, \bibinfo {author} {\bibfnamefont {T.}~\bibnamefont {Kampschulte}}, \bibinfo {author} {\bibfnamefont {M.}~\bibnamefont {Korppi}}, \bibinfo {author} {\bibfnamefont {M.~T.}\ \bibnamefont {Rakher}},\ and\ \bibinfo {author} {\bibfnamefont {P.}~\bibnamefont {Treutlein}},\ }\bibfield  {title} {\bibinfo {title} {Sympathetic cooling of a membrane oscillator in a hybrid mechanical--atomic system},\ }\href {https://doi.org/10.1038/nnano.2014.278} {\bibfield  {journal} {\bibinfo  {journal} {Nat. Nanotechnol.}\ }\textbf {\bibinfo {volume} {10}},\ \bibinfo {pages} {55} (\bibinfo {year} {2015})}\BibitemShut {NoStop}%
\bibitem [{\citenamefont {Guo}\ \emph {et~al.}(2014)\citenamefont {Guo}, \citenamefont {Li}, \citenamefont {Nie},\ and\ \citenamefont {Li}}]{2cavity1}%
  \BibitemOpen
  \bibfield  {author} {\bibinfo {author} {\bibfnamefont {Y.}~\bibnamefont {Guo}}, \bibinfo {author} {\bibfnamefont {K.}~\bibnamefont {Li}}, \bibinfo {author} {\bibfnamefont {W.}~\bibnamefont {Nie}},\ and\ \bibinfo {author} {\bibfnamefont {Y.}~\bibnamefont {Li}},\ }\bibfield  {title} {\bibinfo {title} {Electromagnetically-induced-transparency-like ground-state cooling in a double-cavity optomechanical system},\ }\href {https://doi.org/10.1103/PhysRevA.90.053841} {\bibfield  {journal} {\bibinfo  {journal} {Phys. Rev. A}\ }\textbf {\bibinfo {volume} {90}},\ \bibinfo {pages} {053841} (\bibinfo {year} {2014})}\BibitemShut {NoStop}%
\bibitem [{\citenamefont {Yang}\ \emph {et~al.}(2019)\citenamefont {Yang}, \citenamefont {Wang}, \citenamefont {Bai}, \citenamefont {Guan}, \citenamefont {Gao}, \citenamefont {Zhu},\ and\ \citenamefont {Wang}}]{2cavity2}%
  \BibitemOpen
  \bibfield  {author} {\bibinfo {author} {\bibfnamefont {J.-Y.}\ \bibnamefont {Yang}}, \bibinfo {author} {\bibfnamefont {D.-Y.}\ \bibnamefont {Wang}}, \bibinfo {author} {\bibfnamefont {C.-H.}\ \bibnamefont {Bai}}, \bibinfo {author} {\bibfnamefont {S.-Y.}\ \bibnamefont {Guan}}, \bibinfo {author} {\bibfnamefont {X.-Y.}\ \bibnamefont {Gao}}, \bibinfo {author} {\bibfnamefont {A.-D.}\ \bibnamefont {Zhu}},\ and\ \bibinfo {author} {\bibfnamefont {H.-F.}\ \bibnamefont {Wang}},\ }\bibfield  {title} {\bibinfo {title} {Ground-state cooling of mechanical oscillator via quadratic optomechanical coupling with two coupled optical cavities},\ }\href {https://doi.org/10.1364/OE.27.022855} {\bibfield  {journal} {\bibinfo  {journal} {Opt. Express}\ }\textbf {\bibinfo {volume} {27}},\ \bibinfo {pages} {22855} (\bibinfo {year} {2019})}\BibitemShut {NoStop}%
\bibitem [{\citenamefont {Mansouri}\ \emph {et~al.}(2022)\citenamefont {Mansouri}, \citenamefont {Rezaie}, \citenamefont {Ranjbar},\ and\ \citenamefont {Daeichian}}]{2cavity3}%
  \BibitemOpen
  \bibfield  {author} {\bibinfo {author} {\bibfnamefont {D.}~\bibnamefont {Mansouri}}, \bibinfo {author} {\bibfnamefont {B.}~\bibnamefont {Rezaie}}, \bibinfo {author} {\bibfnamefont {A.}~\bibnamefont {Ranjbar}},\ and\ \bibinfo {author} {\bibfnamefont {A.}~\bibnamefont {Daeichian}},\ }\bibfield  {title} {\bibinfo {title} {Cavity-assisted coherent feedback cooling of a mechanical resonator to the ground-state in the unresolved sideband regime},\ }\href {https://doi.org/10.1088/1361-6455/ac7d27} {\bibfield  {journal} {\bibinfo  {journal} {J. Phys. B}\ }\textbf {\bibinfo {volume} {55}},\ \bibinfo {pages} {165501} (\bibinfo {year} {2022})}\BibitemShut {NoStop}%
\bibitem [{\citenamefont {Liu}\ \emph {et~al.}(2015)\citenamefont {Liu}, \citenamefont {Xiao}, \citenamefont {Luan}, \citenamefont {Gong},\ and\ \citenamefont {Wong}}]{YCLiu2015pra}%
  \BibitemOpen
  \bibfield  {author} {\bibinfo {author} {\bibfnamefont {Y.-C.}\ \bibnamefont {Liu}}, \bibinfo {author} {\bibfnamefont {Y.-F.}\ \bibnamefont {Xiao}}, \bibinfo {author} {\bibfnamefont {X.}~\bibnamefont {Luan}}, \bibinfo {author} {\bibfnamefont {Q.}~\bibnamefont {Gong}},\ and\ \bibinfo {author} {\bibfnamefont {C.~W.}\ \bibnamefont {Wong}},\ }\bibfield  {title} {\bibinfo {title} {Coupled cavities for motional ground-state cooling and strong optomechanical coupling},\ }\href {https://doi.org/10.1103/PhysRevA.91.033818} {\bibfield  {journal} {\bibinfo  {journal} {Phys. Rev. A}\ }\textbf {\bibinfo {volume} {91}},\ \bibinfo {pages} {033818} (\bibinfo {year} {2015})}\BibitemShut {NoStop}%
\bibitem [{\citenamefont {Dantan}\ \emph {et~al.}(2014)\citenamefont {Dantan}, \citenamefont {Nair}, \citenamefont {Pupillo},\ and\ \citenamefont {Genes}}]{GenesDoped2014}%
  \BibitemOpen
  \bibfield  {author} {\bibinfo {author} {\bibfnamefont {A.}~\bibnamefont {Dantan}}, \bibinfo {author} {\bibfnamefont {B.}~\bibnamefont {Nair}}, \bibinfo {author} {\bibfnamefont {G.}~\bibnamefont {Pupillo}},\ and\ \bibinfo {author} {\bibfnamefont {C.}~\bibnamefont {Genes}},\ }\bibfield  {title} {\bibinfo {title} {Hybrid cavity mechanics with doped systems},\ }\href {https://doi.org/10.1103/PhysRevA.90.033820} {\bibfield  {journal} {\bibinfo  {journal} {Phys. Rev. A}\ }\textbf {\bibinfo {volume} {90}},\ \bibinfo {pages} {033820} (\bibinfo {year} {2014})}\BibitemShut {NoStop}%
\bibitem [{\citenamefont {{\v{C}}ernot{\'\i}k}\ \emph {et~al.}(2019{\natexlab{a}})\citenamefont {{\v{C}}ernot{\'\i}k}, \citenamefont {Dantan},\ and\ \citenamefont {Genes}}]{FanoM}%
  \BibitemOpen
  \bibfield  {author} {\bibinfo {author} {\bibfnamefont {O.}~\bibnamefont {{\v{C}}ernot{\'\i}k}}, \bibinfo {author} {\bibfnamefont {A.}~\bibnamefont {Dantan}},\ and\ \bibinfo {author} {\bibfnamefont {C.}~\bibnamefont {Genes}},\ }\bibfield  {title} {\bibinfo {title} {Cavity quantum electrodynamics with frequency-dependent reflectors},\ }\href {https://doi.org/10.1103/PhysRevLett.122.243601} {\bibfield  {journal} {\bibinfo  {journal} {Phys. Rev. Lett.}\ }\textbf {\bibinfo {volume} {122}},\ \bibinfo {pages} {243601} (\bibinfo {year} {2019}{\natexlab{a}})}\BibitemShut {NoStop}%
\bibitem [{\citenamefont {{\v{C}}ernot{\'\i}k}\ \emph {et~al.}(2019{\natexlab{b}})\citenamefont {{\v{C}}ernot{\'\i}k}, \citenamefont {Genes},\ and\ \citenamefont {Dantan}}]{GenesQST}%
  \BibitemOpen
  \bibfield  {author} {\bibinfo {author} {\bibfnamefont {O.}~\bibnamefont {{\v{C}}ernot{\'\i}k}}, \bibinfo {author} {\bibfnamefont {C.}~\bibnamefont {Genes}},\ and\ \bibinfo {author} {\bibfnamefont {A.}~\bibnamefont {Dantan}},\ }\bibfield  {title} {\bibinfo {title} {Interference effects in hybrid cavity optomechanics},\ }\href {https://doi.org/10.1088/2058-9565/aaf5a6} {\bibfield  {journal} {\bibinfo  {journal} {Quantum Sci. Technol.}\ }\textbf {\bibinfo {volume} {4}},\ \bibinfo {pages} {024002} (\bibinfo {year} {2019}{\natexlab{b}})}\BibitemShut {NoStop}%
\bibitem [{\citenamefont {Monsel}\ \emph {et~al.}(2021)\citenamefont {Monsel}, \citenamefont {Dashti}, \citenamefont {Manjeshwar}, \citenamefont {Eriksson}, \citenamefont {Ernbrink}, \citenamefont {Olsson}, \citenamefont {Torneus}, \citenamefont {Wieczorek},\ and\ \citenamefont {Splettstoesser}}]{Juliette2021pra}%
  \BibitemOpen
  \bibfield  {author} {\bibinfo {author} {\bibfnamefont {J.}~\bibnamefont {Monsel}}, \bibinfo {author} {\bibfnamefont {N.}~\bibnamefont {Dashti}}, \bibinfo {author} {\bibfnamefont {S.~K.}\ \bibnamefont {Manjeshwar}}, \bibinfo {author} {\bibfnamefont {J.}~\bibnamefont {Eriksson}}, \bibinfo {author} {\bibfnamefont {H.}~\bibnamefont {Ernbrink}}, \bibinfo {author} {\bibfnamefont {E.}~\bibnamefont {Olsson}}, \bibinfo {author} {\bibfnamefont {E.}~\bibnamefont {Torneus}}, \bibinfo {author} {\bibfnamefont {W.}~\bibnamefont {Wieczorek}},\ and\ \bibinfo {author} {\bibfnamefont {J.}~\bibnamefont {Splettstoesser}},\ }\bibfield  {title} {\bibinfo {title} {Optomechanical cooling with coherent and squeezed light: The thermodynamic cost of opening the heat valve},\ }\href {https://doi.org/10.1103/PhysRevA.103.063519} {\bibfield  {journal} {\bibinfo  {journal} {Phys. Rev. A}\ }\textbf {\bibinfo {volume} {103}},\ \bibinfo {pages} {063519} (\bibinfo {year} {2021})}\BibitemShut {NoStop}%
\bibitem [{\citenamefont {Manjeshwar}\ \emph {et~al.}(2023)\citenamefont {Manjeshwar}, \citenamefont {Ciers}, \citenamefont {Monsel}, \citenamefont {Pfeifer}, \citenamefont {Peralle}, \citenamefont {Wang}, \citenamefont {Tassin},\ and\ \citenamefont {Wieczorek}}]{WWoe2023}%
  \BibitemOpen
  \bibfield  {author} {\bibinfo {author} {\bibfnamefont {S.~K.}\ \bibnamefont {Manjeshwar}}, \bibinfo {author} {\bibfnamefont {A.}~\bibnamefont {Ciers}}, \bibinfo {author} {\bibfnamefont {J.}~\bibnamefont {Monsel}}, \bibinfo {author} {\bibfnamefont {H.}~\bibnamefont {Pfeifer}}, \bibinfo {author} {\bibfnamefont {C.}~\bibnamefont {Peralle}}, \bibinfo {author} {\bibfnamefont {S.~M.}\ \bibnamefont {Wang}}, \bibinfo {author} {\bibfnamefont {P.}~\bibnamefont {Tassin}},\ and\ \bibinfo {author} {\bibfnamefont {W.}~\bibnamefont {Wieczorek}},\ }\bibfield  {title} {\bibinfo {title} {Integrated microcavity optomechanics with a suspended photonic crystal mirror above a distributed bragg reflector},\ }\href {https://doi.org/10.1364/OE.496447} {\bibfield  {journal} {\bibinfo  {journal} {Opt. Express}\ }\textbf {\bibinfo {volume} {31}},\ \bibinfo {pages} {30212} (\bibinfo {year} {2023})}\BibitemShut {NoStop}%
\bibitem [{\citenamefont {Mitra}\ \emph {et~al.}(2024)\citenamefont {Mitra}, \citenamefont {Singh}, \citenamefont {Darki}, \citenamefont {Madsen},\ and\ \citenamefont {Dantan}}]{Mitra2024Apr}%
  \BibitemOpen
  \bibfield  {author} {\bibinfo {author} {\bibfnamefont {T.}~\bibnamefont {Mitra}}, \bibinfo {author} {\bibfnamefont {G.}~\bibnamefont {Singh}}, \bibinfo {author} {\bibfnamefont {A.~A.}\ \bibnamefont {Darki}}, \bibinfo {author} {\bibfnamefont {S.~P.}\ \bibnamefont {Madsen}},\ and\ \bibinfo {author} {\bibfnamefont {A.}~\bibnamefont {Dantan}},\ }\bibfield  {title} {\bibinfo {title} {{Narrow-linewidth Fano microcavities with resonant subwavelength grating mirror}},\ }\href {https://doi.org/10.1364/OE.521329} {\bibfield  {journal} {\bibinfo  {journal} {Opt. Express}\ }\textbf {\bibinfo {volume} {32}},\ \bibinfo {pages} {15667} (\bibinfo {year} {2024})}\BibitemShut {NoStop}%
\bibitem [{\citenamefont {Monsel}\ \emph {et~al.}(2024)\citenamefont {Monsel}, \citenamefont {Ciers}, \citenamefont {Manjeshwar}, \citenamefont {Wieczorek},\ and\ \citenamefont {Splettstoesser}}]{Juliette2023arxiv}%
  \BibitemOpen
  \bibfield  {author} {\bibinfo {author} {\bibfnamefont {J.}~\bibnamefont {Monsel}}, \bibinfo {author} {\bibfnamefont {A.}~\bibnamefont {Ciers}}, \bibinfo {author} {\bibfnamefont {S.~K.}\ \bibnamefont {Manjeshwar}}, \bibinfo {author} {\bibfnamefont {W.}~\bibnamefont {Wieczorek}},\ and\ \bibinfo {author} {\bibfnamefont {J.}~\bibnamefont {Splettstoesser}},\ }\bibfield  {title} {\bibinfo {title} {Dissipative and dispersive cavity optomechanics with a frequency-dependent mirror},\ }\href {https://doi.org/10.1103/PhysRevA.109.043532} {\bibfield  {journal} {\bibinfo  {journal} {Phys. Rev. A}\ }\textbf {\bibinfo {volume} {109}},\ \bibinfo {pages} {043532} (\bibinfo {year} {2024})}\BibitemShut {NoStop}%
\bibitem [{\citenamefont {Rabl}(2011)}]{SVCR1}%
  \BibitemOpen
  \bibfield  {author} {\bibinfo {author} {\bibfnamefont {P.}~\bibnamefont {Rabl}},\ }\bibfield  {title} {\bibinfo {title} {Photon blockade effect in optomechanical systems},\ }\href {https://doi.org/10.1103/PhysRevLett.107.063601} {\bibfield  {journal} {\bibinfo  {journal} {Phys. Rev. Lett.}\ }\textbf {\bibinfo {volume} {107}},\ \bibinfo {pages} {063601} (\bibinfo {year} {2011})}\BibitemShut {NoStop}%
\bibitem [{\citenamefont {Nunnenkamp}\ \emph {et~al.}(2011)\citenamefont {Nunnenkamp}, \citenamefont {B\o{}rkje},\ and\ \citenamefont {Girvin}}]{SVCR2}%
  \BibitemOpen
  \bibfield  {author} {\bibinfo {author} {\bibfnamefont {A.}~\bibnamefont {Nunnenkamp}}, \bibinfo {author} {\bibfnamefont {K.}~\bibnamefont {B\o{}rkje}},\ and\ \bibinfo {author} {\bibfnamefont {S.~M.}\ \bibnamefont {Girvin}},\ }\bibfield  {title} {\bibinfo {title} {Single-photon optomechanics},\ }\href {https://doi.org/10.1103/PhysRevLett.107.063602} {\bibfield  {journal} {\bibinfo  {journal} {Phys. Rev. Lett.}\ }\textbf {\bibinfo {volume} {107}},\ \bibinfo {pages} {063602} (\bibinfo {year} {2011})}\BibitemShut {NoStop}%
\bibitem [{\citenamefont {Wise}\ \emph {et~al.}(2024)\citenamefont {Wise}, \citenamefont {Dutreix},\ and\ \citenamefont {Pistolesi}}]{Wise2024May}%
  \BibitemOpen
  \bibfield  {author} {\bibinfo {author} {\bibfnamefont {J.~L.}\ \bibnamefont {Wise}}, \bibinfo {author} {\bibfnamefont {C.}~\bibnamefont {Dutreix}},\ and\ \bibinfo {author} {\bibfnamefont {F.}~\bibnamefont {Pistolesi}},\ }\bibfield  {title} {\bibinfo {title} {Nonclassical mechanical states in cavity optomechanics in the single-photon strong-coupling regime},\ }\href {https://doi.org/10.1103/PhysRevA.109.L051501} {\bibfield  {journal} {\bibinfo  {journal} {Phys. Rev. A}\ }\textbf {\bibinfo {volume} {109}},\ \bibinfo {pages} {L051501} (\bibinfo {year} {2024})}\BibitemShut {NoStop}%
\bibitem [{Note1()}]{Note1}%
  \BibitemOpen
  \bibinfo {note} {In order to access well-defined steady-state mean values, we examine the stability of the system by using the Routh-Hurwitz criterion~\cite {DeJesus1987Jun}. We indicate the unstable regime by white areas in all density plots of this paper.}\BibitemShut {Stop}%
\bibitem [{Note2()}]{Note2}%
  \BibitemOpen
  \bibinfo {note} {This approximation is known to give a spurious term in the phase fluctuation spectrum \cite {BrownianN}, but it does not impact the quantity we are interested in, namely the steady-state phonon number, as discussed, e.g., in Appendix of a previous work \cite {Juliette2021pra}}\BibitemShut {NoStop}%
\bibitem [{\citenamefont {Thompson}\ \emph {et~al.}(2008)\citenamefont {Thompson}, \citenamefont {Zwickl}, \citenamefont {Jayich}, \citenamefont {Marquardt}, \citenamefont {Girvin},\ and\ \citenamefont {Harris}}]{Harris2008nature}%
  \BibitemOpen
  \bibfield  {author} {\bibinfo {author} {\bibfnamefont {J.}~\bibnamefont {Thompson}}, \bibinfo {author} {\bibfnamefont {B.}~\bibnamefont {Zwickl}}, \bibinfo {author} {\bibfnamefont {A.}~\bibnamefont {Jayich}}, \bibinfo {author} {\bibfnamefont {F.}~\bibnamefont {Marquardt}}, \bibinfo {author} {\bibfnamefont {S.}~\bibnamefont {Girvin}},\ and\ \bibinfo {author} {\bibfnamefont {J.}~\bibnamefont {Harris}},\ }\bibfield  {title} {\bibinfo {title} {Strong dispersive coupling of a high-finesse cavity to a micromechanical membrane},\ }\href {https://doi.org/10.1038/nature06715} {\bibfield  {journal} {\bibinfo  {journal} {Nature}\ }\textbf {\bibinfo {volume} {452}},\ \bibinfo {pages} {72} (\bibinfo {year} {2008})}\BibitemShut {NoStop}%
\bibitem [{Note3()}]{Note3}%
  \BibitemOpen
  \bibinfo {note} {Actually, the input noise $a_{\protect \text {bot},1}^{\protect \text {in}}$ is the superposition of both the input field coming from the ``bottom environment'' and the output field coming from the Fano mirror, so that it also satisfies the canonical commutation relation.}\BibitemShut {Stop}%
\bibitem [{\citenamefont {Vojna}\ \emph {et~al.}(2019)\citenamefont {Vojna}, \citenamefont {Slez{\ifmmode\acute{a}\else\'{a}\fi}k}, \citenamefont {Lucianetti},\ and\ \citenamefont {Mocek}}]{Verdet2019}%
  \BibitemOpen
  \bibfield  {author} {\bibinfo {author} {\bibfnamefont {D.}~\bibnamefont {Vojna}}, \bibinfo {author} {\bibfnamefont {O.}~\bibnamefont {Slez{\ifmmode\acute{a}\else\'{a}\fi}k}}, \bibinfo {author} {\bibfnamefont {A.}~\bibnamefont {Lucianetti}},\ and\ \bibinfo {author} {\bibfnamefont {T.}~\bibnamefont {Mocek}},\ }\bibfield  {title} {\bibinfo {title} {{Verdet Constant of Magneto-Active Materials Developed for High-Power Faraday Devices}},\ }\href {https://doi.org/10.3390/app9153160} {\bibfield  {journal} {\bibinfo  {journal} {Appl. Sci.}\ }\textbf {\bibinfo {volume} {9}},\ \bibinfo {pages} {3160} (\bibinfo {year} {2019})}\BibitemShut {NoStop}%
\bibitem [{\citenamefont {Genes}\ \emph {et~al.}(2008)\citenamefont {Genes}, \citenamefont {Vitali}, \citenamefont {Tombesi}, \citenamefont {Gigan},\ and\ \citenamefont {Aspelmeyer}}]{Genes2008pra}%
  \BibitemOpen
  \bibfield  {author} {\bibinfo {author} {\bibfnamefont {C.}~\bibnamefont {Genes}}, \bibinfo {author} {\bibfnamefont {D.}~\bibnamefont {Vitali}}, \bibinfo {author} {\bibfnamefont {P.}~\bibnamefont {Tombesi}}, \bibinfo {author} {\bibfnamefont {S.}~\bibnamefont {Gigan}},\ and\ \bibinfo {author} {\bibfnamefont {M.}~\bibnamefont {Aspelmeyer}},\ }\bibfield  {title} {\bibinfo {title} {Ground-state cooling of a micromechanical oscillator: Comparing cold damping and cavity-assisted cooling schemes},\ }\href {https://doi.org/10.1103/PhysRevA.77.033804} {\bibfield  {journal} {\bibinfo  {journal} {Phys. Rev. A}\ }\textbf {\bibinfo {volume} {77}},\ \bibinfo {pages} {033804} (\bibinfo {year} {2008})}\BibitemShut {NoStop}%
\bibitem [{\citenamefont {Miroshnichenko}\ \emph {et~al.}(2010)\citenamefont {Miroshnichenko}, \citenamefont {Flach},\ and\ \citenamefont {Kivshar}}]{directcouple}%
  \BibitemOpen
  \bibfield  {author} {\bibinfo {author} {\bibfnamefont {A.~E.}\ \bibnamefont {Miroshnichenko}}, \bibinfo {author} {\bibfnamefont {S.}~\bibnamefont {Flach}},\ and\ \bibinfo {author} {\bibfnamefont {Y.~S.}\ \bibnamefont {Kivshar}},\ }\bibfield  {title} {\bibinfo {title} {Fano resonances in nanoscale structures},\ }\href {https://doi.org/10.1103/RevModPhys.82.2257} {\bibfield  {journal} {\bibinfo  {journal} {Rev. Mod. Phys.}\ }\textbf {\bibinfo {volume} {82}},\ \bibinfo {pages} {2257} (\bibinfo {year} {2010})}\BibitemShut {NoStop}%
\bibitem [{\citenamefont {Xuereb}\ \emph {et~al.}(2009)\citenamefont {Xuereb}, \citenamefont {Horak},\ and\ \citenamefont {Freegarde}}]{Xuereb2009}%
  \BibitemOpen
  \bibfield  {author} {\bibinfo {author} {\bibfnamefont {A.}~\bibnamefont {Xuereb}}, \bibinfo {author} {\bibfnamefont {P.}~\bibnamefont {Horak}},\ and\ \bibinfo {author} {\bibfnamefont {T.}~\bibnamefont {Freegarde}},\ }\bibfield  {title} {\bibinfo {title} {Atom cooling using the dipole force of a single retroflected laser beam},\ }\href {https://doi.org/10.1103/PhysRevA.80.013836} {\bibfield  {journal} {\bibinfo  {journal} {Phys. Rev. A}\ }\textbf {\bibinfo {volume} {80}},\ \bibinfo {pages} {013836} (\bibinfo {year} {2009})}\BibitemShut {NoStop}%
\bibitem [{\citenamefont {Tufarelli}\ \emph {et~al.}(2013)\citenamefont {Tufarelli}, \citenamefont {Ciccarello},\ and\ \citenamefont {Kim}}]{FCmirror}%
  \BibitemOpen
  \bibfield  {author} {\bibinfo {author} {\bibfnamefont {T.}~\bibnamefont {Tufarelli}}, \bibinfo {author} {\bibfnamefont {F.}~\bibnamefont {Ciccarello}},\ and\ \bibinfo {author} {\bibfnamefont {M.~S.}\ \bibnamefont {Kim}},\ }\bibfield  {title} {\bibinfo {title} {Dynamics of spontaneous emission in a single-end photonic waveguide},\ }\href {https://doi.org/10.1103/PhysRevA.87.013820} {\bibfield  {journal} {\bibinfo  {journal} {Phys. Rev. A}\ }\textbf {\bibinfo {volume} {87}},\ \bibinfo {pages} {013820} (\bibinfo {year} {2013})}\BibitemShut {NoStop}%
\bibitem [{\citenamefont {Lechner}\ \emph {et~al.}(2023)\citenamefont {Lechner}, \citenamefont {Pennetta}, \citenamefont {Blaha}, \citenamefont {Schneeweiss}, \citenamefont {Rauschenbeutel},\ and\ \citenamefont {Volz}}]{CWtransition}%
  \BibitemOpen
  \bibfield  {author} {\bibinfo {author} {\bibfnamefont {D.}~\bibnamefont {Lechner}}, \bibinfo {author} {\bibfnamefont {R.}~\bibnamefont {Pennetta}}, \bibinfo {author} {\bibfnamefont {M.}~\bibnamefont {Blaha}}, \bibinfo {author} {\bibfnamefont {P.}~\bibnamefont {Schneeweiss}}, \bibinfo {author} {\bibfnamefont {A.}~\bibnamefont {Rauschenbeutel}},\ and\ \bibinfo {author} {\bibfnamefont {J.}~\bibnamefont {Volz}},\ }\bibfield  {title} {\bibinfo {title} {Light-matter interaction at the transition between cavity and waveguide qed},\ }\href {https://doi.org/10.1103/PhysRevLett.131.103603} {\bibfield  {journal} {\bibinfo  {journal} {Phys. Rev. Lett.}\ }\textbf {\bibinfo {volume} {131}},\ \bibinfo {pages} {103603} (\bibinfo {year} {2023})}\BibitemShut {NoStop}%
\bibitem [{\citenamefont {Kockum}(2021)}]{fiveyear}%
  \BibitemOpen
  \bibfield  {author} {\bibinfo {author} {\bibfnamefont {A.~F.}\ \bibnamefont {Kockum}},\ }\bibfield  {title} {\bibinfo {title} {Quantum {O}ptics with {G}iant {A}toms---the {F}irst {F}ive {Y}ears},\ }in\ \href@noop {} {\emph {\bibinfo {booktitle} {International Symposium on Mathematics, Quantum Theory, and Cryptography}}},\ \bibinfo {editor} {edited by\ \bibinfo {editor} {\bibfnamefont {T.}~\bibnamefont {Takagi}}, \bibinfo {editor} {\bibfnamefont {M.}~\bibnamefont {Wakayama}}, \bibinfo {editor} {\bibfnamefont {K.}~\bibnamefont {Tanaka}}, \bibinfo {editor} {\bibfnamefont {N.}~\bibnamefont {Kunihiro}}, \bibinfo {editor} {\bibfnamefont {K.}~\bibnamefont {Kimoto}},\ and\ \bibinfo {editor} {\bibfnamefont {Y.}~\bibnamefont {Ikematsu}}}\ (\bibinfo  {publisher} {Springer Singapore},\ \bibinfo {address} {Singapore},\ \bibinfo {year} {2021})\ pp.\ \bibinfo {pages} {125--146}\BibitemShut {NoStop}%
\bibitem [{\citenamefont {DeJesus}\ and\ \citenamefont {Kaufman}(1987)}]{DeJesus1987Jun}%
  \BibitemOpen
  \bibfield  {author} {\bibinfo {author} {\bibfnamefont {E.~X.}\ \bibnamefont {DeJesus}}\ and\ \bibinfo {author} {\bibfnamefont {C.}~\bibnamefont {Kaufman}},\ }\bibfield  {title} {\bibinfo {title} {Routh-{{Hurwitz}} criterion in the examination of eigenvalues of a system of nonlinear ordinary differential equations},\ }\href {https://doi.org/10.1103/PhysRevA.35.5288} {\bibfield  {journal} {\bibinfo  {journal} {Phys. Rev. A}\ }\textbf {\bibinfo {volume} {35}},\ \bibinfo {pages} {5288} (\bibinfo {year} {1987})}\BibitemShut {NoStop}%
\bibitem [{\citenamefont {Giovannetti}\ and\ \citenamefont {Vitali}(2001)}]{BrownianN}%
  \BibitemOpen
  \bibfield  {author} {\bibinfo {author} {\bibfnamefont {V.}~\bibnamefont {Giovannetti}}\ and\ \bibinfo {author} {\bibfnamefont {D.}~\bibnamefont {Vitali}},\ }\bibfield  {title} {\bibinfo {title} {Phase-noise measurement in a cavity with a movable mirror undergoing quantum brownian motion},\ }\href {https://doi.org/10.1103/PhysRevA.63.023812} {\bibfield  {journal} {\bibinfo  {journal} {Phys. Rev. A}\ }\textbf {\bibinfo {volume} {63}},\ \bibinfo {pages} {023812} (\bibinfo {year} {2001})}\BibitemShut {NoStop}%
\end{thebibliography}%

\end{document}